\documentclass[twoside,11pt]{article}
\usepackage{graphicx}
\usepackage{bm}
\usepackage{float}
\usepackage[round]{natbib}
\usepackage{booktabs}
\usepackage{lipsum}
\usepackage{cite}
\usepackage{appendix}
\usepackage[margin=1in]{geometry}
\usepackage[usenames]{color}
\usepackage{amsfonts, amssymb, amsmath, amsthm, multicol}
\usepackage[colorlinks=true, pdfstartview=FitV, linkcolor=blue,
citecolor=blue, urlcolor=blue]{hyperref}
\usepackage[usenames]{color}
\definecolor{Red}{rgb}{1,0.05,0}
\definecolor{Grn}{rgb}{0.1,0.7,0.1}
\definecolor{Blu}{rgb}{0.1,0.1,0.6}
\definecolor{Org}{rgb}{1,0.45,0}
\definecolor{Vio}{rgb}{0.6578,0,0.9478}
\definecolor{Mag}{rgb}{1,0.2,0.3}
\usepackage{authblk}
\usepackage{booktabs}
\usepackage{caption}
\usepackage{soul}
\usepackage[dvipsnames]{xcolor}

\newcommand{\coeff}{\mathbf{c}}

\author[1]{\textcolor{black}{Lanyu Li}}
\author[1]{\textcolor{black}{Jeffrey  McClure}}
\author[2]{\textcolor{black}{Grady B. Wright}}

\author[3]{\textcolor{black}{Jared P. Whitehead}}
\author[4]{\textcolor{black}{Jin Wang}}
\author[1]{\textcolor{black}{Zhao Pan}\thanks{To whom correspondence may be addressed: zhao.pan@uwaterloo.ca}}
\affil[1]{University of Waterloo, Dept. of Mechanical and Mechatronics Engineering, Waterloo, ON, Canada}
\affil[2]{Boise State University, Dept. of Mathematics, Boise, ID, USA}
\affil[3]{Brigham Young University, Mathematics Department, Provo, UT, USA}
\affil[4]{Independent researcher, Seattle, WA, USA}

\date{\today}

\title{Error propagation of direct pressure gradient integration and a Helmholtz-Hodge decomposition based pressure field reconstruction method for image velocimetry}

\begin{document}

\maketitle

\begin{abstract}
    Recovering pressure fields from image velocimetry measurements has two general strategies: i) directly integrating the pressure gradients from the momentum equation and ii)~solving or enforcing the pressure Poisson equation (divergence of the pressure gradients). 
    In this work, we analyze the error propagation of the former strategy and provide some practical insights. 
    For example, we establish the error scaling laws for the Pressure Gradient Integration (PGI) and the Pressure Poisson Equation (PPE).
    We explain why applying the Helmholtz-Hodge Decomposition (HHD) could significantly reduce the error propagation for the PGI.
    We also propose to use a novel HHD-based pressure field reconstruction strategy that offers the following advantages \textcolor{black}{or features}: 
    i) effective processing of noisy scattered or structured image velocimetry data on a complex domain; 
    ii) using Radial Basis Functions (RBFs) with divergence/curl-free kernels to provide divergence-free correction to the velocity fields \textcolor{black}{for incompressible flows} and curl-free correction for pressure gradients; and
    iii) \textcolor{black}{enforcing divergence/curl-free constraints without using Lagrangian multipliers.}
    Complete elimination of divergence-free bias in measured pressure gradient and curl-free bias in the measured velocity field results in superior accuracy.
    Synthetic velocimetry data based on exact solutions and high-fidelity simulations are used to validate the analysis as well as demonstrate the flexibility and effectiveness of the RBF-HHD solver. 

\end{abstract}

\section{Introduction}
\label{Sect. Intro}

Reconstructing pressure fields by velocity fields measured from image velocimetry --- Particle Image Velocimetry (PIV) or Particle Tracking Velocimetry (PTV), for example---is an attractive strategy for noninvasive pressure field measurement. 
This idea can be traced back to \citet{schwabe1935druckermittlung,gurka1999computation}, and the research field has been significantly advanced in recent years \citep{van2013piv,van2017comparative,sperotto2022meshless}.
In general, there are two strategies to reconstruct the pressure fields: i) integrating the pressure gradient \textcolor{black}{(e.g., from the momentum equations of an incompressible flow):}
\begin{equation}
\label{eq: grad p}
    \nabla p = \bm{g}(\bm{u}) = - \rho  \left( \frac{\partial \bm{u} }{\partial t} + (\bm{u} \cdot \nabla )\bm{u} - \nu \nabla^2 \bm{u} \right),
\end{equation}
where $p$ is the pressure field to be recovered, $\bm{u}$ is the velocity field measured by the image velocimetry experiments, and $\bm{g}(\bm{u})$ is a function of $\bm u$ derived from the momentum equations.  
We call this strategy direct Pressure Gradient Integration (PGI) in the current work. 
The second option is ii)~solving the Pressure Poisson Equation (PPE):
\begin{equation}
\label{eq: PPE}
    \nabla^2 p = f(\bm{u}) = - \rho  \nabla \cdot \left( \frac{\partial \bm{u} }{\partial t} + (\bm{u} \cdot \nabla )\bm{u} - \nu \nabla^2 \bm{u} \right),
\end{equation}
which is equivalent to integrating twice for $p$ with proper boundary conditions. 
Alternatively, a data assimilation or machine learning based method often enforces \eqref{eq: grad p} and/or \eqref{eq: PPE} by constructing constraints or loss functions with them, and we may interpret these methods as indirect integration strategies. 
A concise review of these methods can be found, for example, in \citet{zhou2023stochastic}. 
In the current study, we focus on analyzing the direct integration of $\nabla p$ or $\nabla^2 p$ for pressure reconstruction.

The PPE is often solved by numerical schemes such as the finite element method \citep{pirnia2020estimating}, the method of fundamental solutions \citep{goushcha2023modified}, and the widely adopted finite difference method \citep{gurka1999computation,ghaemi2012piv,de2012instantaneous,van2019pressure}. 
Recent advancement in Lagrangian particle tracking motivates the use of mesh-free methods to solve the PPE
\citep{sperotto2022meshless}.
One fundamental advantage of solving the Poisson equation is that given \textcolor{black}{certain} data $f(\bm u)$ and proper boundary conditions, the reconstructed pressure field exists and is unique (up to a constant reference pressure).
\textcolor{black}{More specifically, even if the data and boundary conditions are corrupted, solving PPE would still produce a deterministically unique, albeit contaminated, pressure field \citep{Pryce2024revisit}.}
However, some benchmarking studies and rigorous error analysis show that solving the PPE could be sensitive to the error in the data 
\citep{charonko2010assessment,Pan2016Error1}.

\textcolor{black}{Directly integrating} \eqref{eq: grad p} can also provide pressure reconstruction.
However, since the pressure gradient is a vector field (e.g., $\nabla p  = (\partial_x p, \partial_y p)$ in 2D), integrating from different starting points in the domain along different paths typically gives conflicting results for the pressure at a fixed location if the velocimetry data are corrupted.
This indicates an ill-posed problem: a (unique) solution may not exist.
Pursuing a unique solution to a such problem is not trivial and some regulation strategies are needed, some of which are summarized below.

\citet{baur1999piv} integrated the pressure gradient starting from the edge of the domain using a spatial-marching strategy. 
The pressure value at an unknown nodal point is computed by a weighted average of the pressure values obtained by integrating from its four immediate (anisotropic) neighbors with known pressure values. 
These four integrals are along short paths from four different directions and give different pressure values for the same location. 
The weighted average provides a regularization strategy attempting to identify unique results.
Because the pressure field $p$ is a scalar potential, and the integration of the pressure gradient ($\nabla p$) must be path-independent, an alternative interpretation of this algorithm is that the Path-Independence Property (PIP) of the integral on the pressure gradient field is enforced \textit{locally}. 
The original implementation of this method starts the integration from one edge\footnote{In theory, the method can start the integration from one point on the boundary or in the domain.} of a rectangular domain and the solution marches towards the other side of the domain. 
This directional algorithm may have some spurious anisotropic properties. 
For example, starting the algorithm from different edges may lead to different pressure reconstructions using the same pressure gradient data. 

An advancement by \citet{liu2006instantaneous} largely resolved the potential anisotropic issues of directly integrating a pressure gradient field. 
A cluster of rays starting from one point on the virtual boundary covering the actual domain is used as the guidelines for the integration. 
Integration of the pressure gradients along these rays\footnote{The integration is only applied within the actual domain, and the integration paths are the ``zig-zag'' routes on the grid crossing the nodal points along the guiding rays.} gives pressure values at the nodal points in the actual domain. 
Rotating the guiding rays by changing the starting point on the virtual boundary generates a new ensemble of integration paths. 
Integration along these paths updates the pressure field which was computed on prior integration paths. 
Iterating this procedure and exhausting the nodal points on the virtual boundary until the pressure field converges gives a `non-biased' pressure field reconstruction. 
The core idea behind this method is to perform a finite ensemble reconstruction of the pressure field on a discrete mesh aimed at satisfying the PIP of the line integral on the gradient of a scalar field.
Compared to \citet{baur1999piv}, the advantage of this algorithm is that \textit{global} enforcement of the PIP is attempted by `exhausting' straight integration paths crossing the domain from `all' directions. 
This idea penned the ``Omini-Directional Integral'' (ODI) method and is later referred to as Circular Virtual Boundary Omni-Directional Integration (CVB-ODI) \citep{liu2016instantaneous} to differentiate it from the more recent development of this family of methods.
It has been shown that when mean-zero random noise is present in the pressure gradient, the ODI can effectively `cancel' the random noise in the data by taking advantage of ensemble averaging \citep{liu2006instantaneous,charonko2010assessment}. 
Despite these advantages, this method can be sensitive to the size and shape of the virtual boundary \citep{liu2016instantaneous,Wang2023private} due to the fact that the geometry of the virtual boundary affects the arrangement of the guiding rays and, in turn, impacts the effective weights for the averaging and the pressure reconstruction: when using a finite-sized virtual boundary, higher weights are assigned to the nodal points close to the boundaries of the domain \textcolor{black}{than to those farther away}.

\citet{liu2016instantaneous} and \citet{wang2019gpu} improved CVB-ODI using rotating parallel rays as the guidelines for integration.
This practice leads to a uniformly weighted average and potentially improves the accuracy relative to the original CVB-ODI.
This algorithm is referred to as the Rotating Parallel Ray Omni-Directional Integration (RPR-ODI) or parallel-line ODI and can be viewed as the CVB-ODI with an infinitely large virtual boundary. 
\citet{wang2019impact,Wang2023private} argues that these ODI iterations may be viewed (and equivalently formulated) as a large-scale weighted least squares regression that seeks a potential field whose interior and boundary values satisfy the PIP as closely as possible, where the weights are indirectly determined by the arrangement of the integral paths.

Following the idea of ODI, several methods have been proposed to reduce the prohibitive computational cost of the conventional ODI.
\citet{dabiri2013algorithm} used low-pass filters and the median of eight integration paths to reconstruct the pressure field. 
\citet{wang2019gpu} used the Graphics Processing Unit (GPU) to accelerate the computation of the ODI and applied it to pressure field reconstruction in 3D. 
\citet{Charonko2022Computational,zigunov2024fast} formulated the ODI as a relaxed iterative algorithm to eliminate the need to explicitly calculate and store each individual line integral to improve the computational efficiency. 
The implicit algorithm can achieve similar or slightly higher accuracy than the traditional ODI methods with computational cost on par with Poisson solvers. 
\textcolor{black}{The following study by the same author \citep{zigunov2024one} suggested that the ODI method is closely related to the Poisson equation. Using the conjugate gradient method to solve the corresponding linear system can yield solutions thousands to millions of times faster than conventional ODI methods on their tests. A rigorous interpretation of \citep{zigunov2024one} can be found in \citep{Pryce2024revisit}, which shows that the solution out of RPR-ODI is exactly a minimal norm least square solution to a Neumann problem of the Poisson equation.}

Aside from computational cost considerations, the quantification of errors and uncertainties is of significant importance in pressure field reconstruction through PPE or PGI methods. 
\citet{Pan2016Error1} analyzed the error propagation from the data field $f = \nabla^2 p$ to the pressure field $p$ for the Poisson equation \eqref{eq: PPE} and provided an estimate (upper bound) for the error in the pressure field. 
These upper bounds can be considered a conservative uncertainty quantification for the PPE-based pressure field reconstruction.
Convergence of ODI was analyzed by \citet{liu2020error}. 
This work claims that the error in the recovered pressure vanishes when the number of integration paths approaches the infinite limit.
However, this conclusion is based on a critical assumption that the error in the pressure gradient must be mean zero random noise. 
If any bias error persists in the data without any corrections, ODI or PPE-based solvers may not be able to approach an unbiased pressure reconstruction. 
In fact, bias errors arising in calibration due to, for example, misalignment and lens distortion, are common in the PIV/PTV measurements \citep{lee2022effect, Raffel2018}, and small bias taking a certain profile could be significantly amplified by the pressure field reconstruction \citep{faiella2021error}.

\textcolor{black}{Unfortunately, the systematic bias typically cannot be removed by averaging; therefore, some physical constraints or prior information may be leveraged to regularize the data and/or the reconstruction.}
For example, the divergence-free property of the incompressible velocity field has been utilized in several correction schemes to reduce both velocity and pressure field errors \citep{de2013minimization,azijli2015solenoidal,wang2017weighted,schiavazzi2017effect}. 
\citet{wang2016irrotation} enforced the curl-free condition of the pressure gradient field preceding their pressure integration scheme. 
The PIP of the pressure integral over the corrected pressure gradient field eliminated the need for computational intensive averaging from many integration paths.
Prior to solving for pressure, \citet{huhn2016fft} penalized high frequencies and non-zero curl of their acceleration field data using smoothing basis splines (B-splines) according to the FlowFit method developed by \citet{gesemann2016noisy}. 
\citet{mcclure2017instantaneous} formulated a divergence-curl system of equations for the pressure gradient error field, which may be recovered successfully if all terms in the formulation are known. 
However, there is no existing method to measure pressure gradient errors on the boundaries of the domain, and certain terms in the pressure gradient divergence must be estimated. 
\citet{lin2023Divergence} started from the acceleration field measured from Lagrangian particle tracking data and solved an optimization problem for the acceleration field constrained to satisfy zero pressure gradient curl and zero viscous term divergence. 
Similarly, \citet{mcclure2019generalized} extended their divergence-curl system for pressure gradients estimated from acceleration fields derived by Lagrangian methods.

Despite demonstrating successful or improved reconstruction based on divergence/curl-free regularization, most of these sophisticated reconstruction methods did not provide corresponding error estimation, which could be intractable depending on the specific algorithm. 
In addition, undertaking trials without understanding the mechanism of error propagation may result in unpredictable outcomes.
This gap motivates the current research: 
there is hope to study the error propagation of the direct PGI similar to what one can do for the PPE.
Such error analysis can be insightful and of practical use: it may provide guidelines and inspire novel pressure field reconstruction methods to control and reduce the error propagation either by PGI or PPE.

\section{Error Estimate for PGI (with Curl-free Regularization)}
\label{sec: Error Estimate for PGI (with Curl-free Regularization)}
In this section, we provide an error estimate of the pressure field reconstructed by PGI.
Let the uncontaminated pressure gradient be
\begin{equation}
    \label{eq: clean grad p}
    {\boldsymbol{g}}= \nabla p  \quad \text{in} ~ \Omega,
\end{equation}
where $p$ is the pressure field with the boundary value of  
\begin{equation}
    \label{eq: clean p on boundary}
    p  =  p_0 \quad \text{on} ~ \partial \Omega. 
\end{equation}
As $p$ is a scalar potential, its gradient $\bm{g}$ is a conservative vector field that is curl-free (i.e., $\nabla \times \bm{g} = \bm{0}
 $). 
A fundamental property of a conservative field is that the line integral from one location in the field to the other is independent of the path of integral, as specified by the gradient theorem.

If $\nabla p$ is corrupted by some  error $\epsilon_{\nabla p}$: 
\begin{equation}
    \label{eq: corrupted grad p}
\tilde{\boldsymbol{g}} =  \nabla p + \epsilon_{\nabla p} \quad \text{in} ~ \Omega, 
\end{equation}
the PIP does not necessarily hold for the contaminated field $\tilde{\boldsymbol{g}}$, \textcolor{black}{as $\epsilon_{\nabla p}$ may have divergence-free component in addition to curl-free, and harmonic parts}. 
However, this issue can be resolved by applying the Helmholtz-Hodge Decomposition (HHD) on the contaminated pressure field.

One of the formulations of the HHD states that a sufficiently smooth vector field $\bm{\xi}$ on a bounded domain $\Omega$ with boundary $\partial \Omega$ can be \textit{uniquely} decomposed in the form
$\bm{\xi} = \nabla \varphi + \bm{r},$
where the vector field $\bm{r}$ is divergence-free; $\varphi$ is a scalar potential and $\nabla \varphi$ is curl-free
\citep{chorin1990mathematical}.
For example, a contaminated pressure gradient field $\tilde{\bm{g}}$ can be decomposed as
\begin{equation}
    \label{eq: corrupted grad p HHD}
\tilde{\bm{g}} =  \nabla p +\epsilon_{\nabla p} = \nabla \varphi + \bm{r}_{\nabla p} \quad   \text{in} ~ \Omega,
\end{equation}
and $\bm{r}_{\nabla p}$ is \textcolor{black}{the divergence-free part of $\epsilon_{\nabla p}$ and is} removable. 
If the conditions 
on the boundary, e.g., $\hat{\bm{n}} \cdot \bm{r}_{\nabla p} = \bm{0}$ are given,  $\hat{\bm{n}} \cdot \nabla \varphi = d_n$ is naturally determined. 
In turn, $\varphi$ is unique up to a constant, including the value of $\varphi$ on the boundary: 
\begin{equation}
    \label{eq: corrupted p on boundary}
    \varphi  =  \varphi_0 \quad \text{on} ~ \partial  \Omega. 
\end{equation}

The significance of HHD in the context of pressure reconstruction is that the resulting gradient field $\nabla \varphi$  \textit{exactly} satisfies the PIP for a pressure gradient field and the divergence-free component in $\tilde{\bm{g}}$ can be \textit{completely} removed. 
This allows consistent reconstruction of $p \approx \varphi$ based on the argument of  $ \nabla p \approx \nabla \varphi$. 
A comprehensive survey by \citet{bhatia2012helmholtz} on HHD, which covers the theoretical aspects, practical applications, and computational techniques, is a valuable resource.

\textcolor{black}{Here, we emphasize the subtle differences between $p$ and $\varphi$.  $\nabla \varphi$ is curl-free, not exactly error-free. Thus, $\varphi$ is an approximation of the true pressure field (i.e., $p \approx \varphi$)}.
The outcome of the HHD ($\nabla \varphi$) only guarantees the curl-free property and thus $\varphi$ is a scalar potential.
\textcolor{black}{The HHD only identifies and removes the divergence-free part in $\tilde{\bm{g}}$ (i.e., $\bm{r}_{\nabla p}$), which should not be there at all. 
The deviation between $\varphi$ and $p$ is the error in the reconstructed pressure field when HHD is used as a regularization method, and this error is associated with both curl-free and harmonic components in $\epsilon_{\nabla p}$, which cannot be identified and removed from corrupted pressure gradients using HHD alone.}

We next investigate the error ($\varphi - p$) in the reconstructed pressure field when the contaminated pressure gradient field is integrated.
Comparing \eqref{eq: clean grad p} and \eqref{eq: corrupted grad p HHD} isolates the error in the pressure gradient
\begin{equation}
    \label{eq: error in field}
\epsilon_{\nabla p}  =  \tilde{\bm{g}} - \bm{g} = \nabla (\varphi-p) + \bm{r}_{\nabla p}   \quad \text{in} ~   \Omega, 
\end{equation}
and the discrepancy on the boundary:
\begin{equation}
    \label{eq: error on boundary}
\epsilon_p  =  \varphi_0 - p _0   \quad \text{on} ~ \partial  \Omega. 
\end{equation}
Computing the norm of both sides of \eqref{eq: error in field} and applying the triangle inequality leads to
\begin{equation}
\begin{split}
    \label{eq: bound for grad D - p}
\| \epsilon_{\nabla p} \|_{L^2(\Omega)} 
&=  \| \nabla  (\varphi-p) + \bm{r}_{\nabla p}  \|_{L^2(\Omega)} \\
&\geq  \| \nabla  (\varphi-p)  \|_{L^2(\Omega)} - \|  \bm{r}_{\nabla p}  \|_{L^2(\Omega)}  \quad \text{in} ~  \Omega. 
\end{split}
\end{equation}
Using the Poincar\'e inequality for the discrepancy between the reconstruction results of the HHD and the true value of the pressure and combining this with \eqref{eq: bound for grad D - p} gives the bound
\begin{equation}
\label{eq: error bound for D - p}
     \| \epsilon_p \|_{L^2(\Omega)}  =      \| \varphi-p \|_{L^2(\Omega)}  \leq C \left( \| \epsilon_{\nabla p} \|_{L^2(\Omega)}  + \| \bm{r}_{\nabla p}  \|_{L^2(\Omega)} \right),
\end{equation}
where $C$ is the Poincar\'e constant.
The value of $C$ is independent of the numerical scheme of the pressure solvers or experimental methods (e.g., PIV or PTV).
If the error on the boundary is considered, the error in the reconstructed pressure field is higher than indicated in \eqref{eq: error bound for D - p} and can be bounded as
\begin{equation}
    \label{eq: error bound ODI}
\| \epsilon_p \|_{L^2(\Omega)}  \leq  C \left( \| \epsilon_{\nabla p} \|_{L^2(\Omega)}  + \| \bm{r}_{\nabla p}  \|_{L^2(\Omega)} \right) + ||\varphi_0 - p_0||_{L^\infty(\partial\Omega)}. 
\end{equation}

\section{Practical Insights from the Error Analysis}
\label{Sec: Insights from the error estimation}

The error estimate in \eqref{eq: error bound for D - p} or \eqref{eq: error bound ODI} can provide insights into engineering practice and is of theoretical interest to the experimental community. 
We will demonstrate a few of the potential insights these bounds provide in the following sections.

\subsection{HHD-based regularization for improved pressure reconstruction}
\label{Sec: HHD-based regularization}

If a proper HHD is performed to regularize the pressure gradient, $\bm{r}_{\nabla p}$ can be removed from  $\tilde{\bm{g}}$, and using $\varphi$ alone to reconstruct the pressure field would reduce the error in the pressure field:
\begin{subequations}
\begin{align}
    \| \epsilon_p \|_{L^2(\Omega)}  &\leq  C  \| \epsilon_{\nabla p} \|_{L^2(\Omega)} \label{eqn:error bound HHD-1} \\
    &\leq  C  \| \epsilon_{\nabla p} \|_{L^2(\Omega)} +  || \varphi_0 - p_0 ||_{L^\infty(\partial \Omega)} \label{eqn:error bound HHD-2}.
\end{align}
\end{subequations}
It is obvious that the estimated error in \eqref{eqn:error bound HHD-1} is potentially lower than that for \eqref{eq: error bound for D - p}, by a difference of $C||\bm{r}||_{L^2{(\Omega})}$ (similar for \eqref{eqn:error bound HHD-2} and \eqref{eq: error bound ODI}). 
The effectiveness of this simple practice will be demonstrated using a Radial Basis Function (RBF) based HHD solver in later sections.
We emphasize that the above analysis and the error estimate are generic and independent of the experimental techniques or the specific implementation of HHD (e.g., some existing works by \citet{wang2016irrotation,wang2017weighted,mcclure2019generalized,lin2023Divergence} \textcolor{black}{pursuing} the PIP by HHD).

In fact, the idea of enforcing the PIP have been pursued from various perspectives to different extents.
Examples include the works by \citet{baur1999piv,liu2006instantaneous,dabiri2013algorithm,wang2016irrotation,wang2017weighted,mcclure2019generalized}, \citet{lin2023Divergence} and \citet{pryce2024simple}.
The goal shared by these methods is to recover the pressure field from a contaminated pressure gradient field by seeking a curl-free pressure gradient field. 
These methods fulfilled the goal with different levels of success. 
Comments on some of these solvers can be found in Sect.~\ref{Sect. Intro}.

\subsection{HHD-regularization as a ``limit'' of ODI}
\label{Sec: HHD as a continuous limit of ODI}
The goal of ODI, an iconic pressure reconstruction method, is to recover a pressure field aiming to satisfy the PIP of the pressure field. 
Invoking the curl-free property of the $\nabla \varphi$ from the HHD, which \textit{exactly} satisfies PIP, we can argue that the limit of applying ODI on a corrupted pressure gradient field is carrying out HHD-based regularization to the same field.
If it is ever possible to find a bound for the error in the pressure computed by ODI, it is expected to be (slightly) higher than that of \eqref{eqn:error bound HHD-2}. 

In addition, the ODI family of methods may present some fundamental features that extend beyond the potentially high computational cost, which, with recent improvements in computational efficiency, may have become a minor concern \citep{wang2019gpu,wang2023green,zigunov2024fast,zigunov2024one}: i) the ODI cannot differentiate curl-free error in the pressure gradient, as the curl-free error also satisfies the PIP. 
This is the same as a common HHD-based method. 
Thus, such errors in the pressure gradient can penetrate through reconstruction, by either the ODI family or HHD-based methods, and contaminate the recovered pressure fields;
ii) the ODI does not necessarily reject any divergence-free components in the pressure gradient (i.e., $\bm{r}_{\nabla p}$), while the divergence-free error could be removed by the HHD-based methods.
Recent advancement of ODI \citep{zigunov2024fast,zigunov2024one} showed that the ODI family of methods has close connections to the PPE. 
Despite that the original motivation of the ODI algorithm is to iteratively converge to a pressure field by enforcing the PIP on corrupted pressure gradients, the simplification by \citet{zigunov2024one} does not show that the curl-free correction intended by ODI is actually enforced.
\textcolor{black}{\citet{prycerevisit,Pryce2024revisit} confirmed this by showing that the ODI is equivalent to pursuing the minimal-norm least square solution to a Neumann problem of PPE.}  
Evidence supporting this argument can also be found in the later sections.


\subsection{Connections between PPE and HHD-regularized PGI}
\label{Sec: Connections between PPE and HHD}

Assuming $\bm{r}_{\nabla p}$ is small (which is often the case for high-quality velocimetry data) or is removed (which is the case when applying HHD), ignoring $\| \bm{r}_{\nabla p}\|_{L^2(\Omega)}$ in \eqref{eq: error bound for D - p} leads to 
\begin{equation}
\label{eq: bound for D-p, no r}
     \| \varphi-p \|_{L^2(\Omega)}  \leq C  \| \epsilon_{\nabla p} \|_{L^2(\Omega)}.  
\end{equation}
More explicitly, this assumption implies only the curl-free part of $\epsilon_{\nabla p}$ is considered for error propagation. 
This means that $\epsilon_{\nabla p} = \nabla \epsilon_{p}$ and $\epsilon_{p} = \varphi - p$. In turn, \eqref{eq: bound for D-p, no r} turns into 
\begin{equation}
    \label{eq: bound for epsilon p, HHD}
        \| \epsilon_p\|_{L^2(\Omega)}  \leq C  \|  \nabla \epsilon_p \|_{L^2(\Omega)}.
\end{equation}

Recalling the error estimate of the PPE, which is based on the analysis of the Poisson equation with respect to $\epsilon_p$ (details can be found in 
\citet{Pan2016Error1}):  
\begin{equation}
    \label{eq: PPE epsilon_p}
  \nabla^2 \epsilon_p  =   \nabla \cdot \nabla \epsilon_p =  \epsilon_f,
\end{equation}
where $\epsilon_f$ is the error in the data of the PPE.
Applying the Poincar\'e inequality to \eqref{eq: PPE epsilon_p} gives
\begin{equation}
    \label{eq: bound grad p, PPE}
    \| \nabla \epsilon_p\|_{L^2(\Omega)}  \leq C  \|  \epsilon_f \|_{L^2(\Omega)},
\end{equation}
and combining \eqref{eq: bound for epsilon p, HHD} and \eqref{eq: bound grad p, PPE} recovers the error bound by directly analyzing the PPE:
\begin{subequations}
\label{eq: bound HHD < PPE}
  \begin{align}
    \| \epsilon_p\|_{L^2(\Omega)}  \leq  \underbrace{C\|\epsilon_{\nabla p}\|_{L^2(\Omega)}}_{\text{bound for HHD-PGI}}  &=  C \|  \nabla \epsilon_p \|_{L^2(\Omega)} \label{eq: bound HHD < PPE -a}\\
    & \leq  \underbrace{C^2 \|  \epsilon_f \|_{L^2(\Omega)}}_{\text{bound for PPE}}. \label{eq: bound HHD < PPE -b}
  \end{align}
\end{subequations}

The error estimate \eqref{eq: bound HHD < PPE} recovers the heuristic that an HHD-regularized PGI solver is expected to outperform a normal Poisson solver (at least in terms of upper bounds).
However, in reality, a direct comparison between these two strategies is not trivial for several reasons. 
First, using the Poincar\'e inequality twice for \eqref{eq: bound HHD < PPE -b} may overestimate the error in the pressure reconstruction more than the result in \eqref{eq: bound HHD < PPE -a}, where the Poincar\'e inequality is applied only once.
Second, the bound for the HHD-based solver scales with the error in the pressure gradient $\|\epsilon_{\nabla p}\|_{L^2(\Omega)}$ but the bound for the PPE-based solver scales with the error in the data field $\|\epsilon_{f}\|_{L^2(\Omega)}$.
The relative value of $\epsilon_{\nabla p}$ and $\epsilon_{f}$ is not immediately obvious, independent of the complexity associated with numerical implementation when evaluating these terms. 
One simple example demonstrating this issue can be found in \citet{Nie2022Error}.
Moreover, the Poincar\'e constants---$C$ and $C^2$ for \eqref{eq: bound HHD < PPE -a} and \eqref{eq: bound HHD < PPE -b}, respectively---depend on the specific properties of the domain such as the geometry and the boundary conditions, and it is not necessarily always true that $C < C^2$.

\subsection{Error scaling for PPE and PGI}
\label{Sec: Error scaling for PPE and PGI}
In this section, we use discrete Fourier analysis to expose the connections in errors between the PPE and PGI.
This study serves a three-pronged purpose: i) investigate the error propagation when the error in the domain is mean-zero random noise, ii) establish the scaling of the error propagation for both PPE and PGI with respect to the size of the domain, the amplitude of the noise, and the resolution of data, as well as iii) corroborate some of the arguments in the previous sections.

We consider pressure field reconstructions using a PPE solver on a two-dimensional (2D) domain of $\Omega = (x, y) \in [0,L]\times [0,L],$  which is discretized into an $N \times N$ grid with grid spacing $h=L/(N-1)$. 
The corresponding error propagation problem is $\nabla^2 \epsilon_p = \epsilon_f$ \citep{Pan2016Error1}, and its 
Fourier transform leads to 
\begin{align}
\label{eq: hat lamda hat}
\hat\epsilon_p[\bm{k}] = - \frac{L^2}{4\pi^2 |\bm{k}|^2} \hat\epsilon_f[k,l] = \lambda_{\bm{k}}^{-1}\hat \epsilon_f[\bm{k}],
\end{align}
where $\hat\epsilon_p[\bm{k}]$ and $\hat \epsilon_f[\bm{k}]$ are the Fourier coefficients for $\epsilon_p$ and $\epsilon_f$, respectively;
and $\bm{k}$ is the wave number.
The coefficients $\lambda_{\bm{k}} = - \frac{L^2}{4\pi^2 |\bm{k}|^2} = - c^2 \frac{L^2}{|\bm{k}|^2}$ will change depending on the geometry of the domain and the boundary conditions available. 
If $\epsilon_f$ is assumed to be a point-wise independent zero mean Gaussian noise with a constant variance $\sigma_f^2$, then truncating \eqref{eq: hat lamda hat} and using orthogonality simplifies everything into a scaling law estimating the error in the pressure field
\begin{equation}
\label{eq: scaling law PPE simple main text}
    \begin{aligned}
    \left \| \epsilon_p \right \|^{\text{PPE}}_{L^2(\Omega)} \sim 
  L^2 \sigma_f N^{-1},
    \end{aligned}
\end{equation}
for a sufficiently large $N$.

For the error in the pressure field reconstructed by the direct PGI, a similar analysis leads to a scaling law estimating the error in the pressure 
\begin{equation}
\label{eq: scaling law dpgi simple main text}
    \begin{aligned}
    \left \| \epsilon_p\right \|^{\text{PGI}}_{L^2(\Omega)} \sim 
  L \sigma_{\bm{g}} N^{-1},
    \end{aligned}
\end{equation}
for a large $N$, where $\sigma^2_{\bm{g}}$ is the variance of the point-wise Gaussian noise in each component of $\tilde{\bm{g}}$.

Noting the $N^{-1}$ term in \eqref{eq: scaling law PPE simple main text} and \eqref{eq: scaling law dpgi simple main text}, the negative power of $N$ recovers the heuristic that high spatial resolution data, corresponding to a large $N$, can lead to an improved signal-to-noise ratio (e.g, $|| \epsilon_p||_{L^2(\Omega)}/\sigma_{f,\bm{g}} \sim N^{-1}$).  
Thus, high-resolution data enable the potential to achieve more accurate reconstruction than the low-resolution ones. 
The $+1$ power for $\sigma_{f,\bm{g}}$ in \eqref{eq: scaling law PPE simple main text} and \eqref{eq: scaling law dpgi simple main text} indicates that both PPE and PGI amplify the random error in their data $\epsilon_f$ and $\epsilon_g$, respectively, in the same way. 
The PPE solver amplifies the error by $L^2$ (see \eqref{eq: scaling law PPE simple main text}), while the error from PGI scales with $L$ (see \eqref{eq: scaling law dpgi simple main text}). 
This is consistent with the Poincar\'e constants of $C^2$ and $C$ in \eqref{eq: bound HHD < PPE}, which is rooted in the fact that solving the PPE requires integrating twice while the PGI needs only a single integration. 

We want to emphasize that the scaling laws with respect to $N$ indicated by \eqref{eq: scaling law PPE simple main text} and \eqref{eq: scaling law dpgi simple main text}, are different from the familiar order of accuracy of a numerical scheme. 
The error estimates in \eqref{eq: scaling law PPE simple main text} and \eqref{eq: scaling law dpgi simple main text} are dominated by the added noise in the data and the effect of the noise masks out the inherent truncation error of the numerical method.
The order of accuracy of the numerical scheme is usually determined by the truncation error, and the corresponding numerical error is not affected by any added error or round-off, but is dictated by the spatial resolution for a given numerical method. 
This subtle difference will be demonstrated in Sect.~\ref{sec: scaling law validation}.

This Fourier analysis can also provide the scaling laws for the variance of the error in the pressure (see Appx.~\ref{sec: scaling laws for PPE and PGI apdx}).  
The variance corresponding to the PGI is consistent with the numerical tests by \citet{wang2023green} and more details can be found in Appx.~\ref{sec: PGI scaling law apdx}.

\section{A Radial Basis Function Based HHD Solver}
\label{sec: RBF-HHD solver}

In this work, we employ a Radial Basis Function (RBF) based HHD solver \citep{fuselier2016high,fuselier2017radial} to reconstruct a pressure field 
for two purposes: i) validate the arguments in Sect.~\ref{Sec: Insights from the error estimation}, and ii) demonstrate the effectiveness of HHD-based solvers in flow reconstructions for PTV data.
\textcolor{black}{We emphasize that our primary objective of this work is to provide an error estimation---specifically, the error bounds and scaling laws detailed in Sect.~\ref{sec: Error Estimate for PGI (with Curl-free Regularization)} and \ref{Sec: Insights from the error estimation}---for direct PGI. 
A by-product of our analysis shows that HHD improves accuracy in pressure reconstruction compared to PGI-based techniques.
To illustrate this in the current section, we use the RBF-HHD solver  \citep{fuselier2017radial}.
For readers who are interested in alternative HHD approaches, we briefly discuss them toward the end of this section.}

The RBF approximants utilize linear combinations of radial kernels that only depend on the Euclidean distance between data points as an independent variable.  Importantly, they do not rely on the data points being \textcolor{black}{gridded} (i.e., they are naturally meshless) so they can be used directly for scattered or structured data measured from PTV or PIV systems in 2D or 3D.
This meshless property together with their excellent approximation properties~\citep{fasshauer2007meshfree} have resulted in RBF methods being widely applied to many areas, including computer graphics \citep{carr2001reconstruction,macedo2011hermite,drake2022implicit}, numerical solutions of differential equations~\citep{FFBook}, machine learning \citep{burkov2019hundred,alpaydin2020introduction}, and data processing for experimental fluid mechanics~\citep{harlander2014orthogonal,sperotto2022meshless,li2024three,ratz2024meshless}. 

The general idea behind the RBF-HHD solver for vector fields in $\mathbb{R}^d$, $d=2$ or $3$, is summarized below
\textcolor{black}{using consistent mathematical notation to that used in \citet{fuselier2016high,fuselier2017radial}.
Readers may refer to their article and Appx.~\ref{sect: solvers hhd} for details. 
We first focus on the $d=3$ case and then generalize to $d=2$.  Consider the \textit{matrix-valued} RBF kernel ${\bf \Phi}:\mathbb{R}^3 \times \mathbb{R}^3 \to \mathbb{R}^3 \times \mathbb{R}^3$ on a bounded domain $\Omega\subset\mathbb{R}^3$ of the form}
\begin{equation}
    {\bf \Phi}(x,y) =  \underbrace{-\textbf{curl}_x\textbf{curl}_y (\phi(\left | x-y \right |) {\bf I})}_{=: \displaystyle {\bf \Phi}^{\text{df}}(x,y)} + \underbrace{\nabla_x (\nabla_y \phi(\left | x-y \right |))^\intercal}_{\displaystyle =: {\bf \Phi}^{\text{cf}}(x,y)},\label{eq:div_curl_kernels3d}
\end{equation}
where $x,y\in\Omega \subset \mathbb{R}^3$, $\left | x-y \right |$ denotes their Euclidean distance, $\bf I$ is the $3$-by-$3$ identity matrix, and the superscript $[\cdot]^\intercal$ represents the transpose.
The operators $\textbf{curl}$ and $\nabla$ are the curl and gradient operators, respectively.
\textcolor{black}{The subscripts (e.g., the $x$ in $\textbf{curl}_x$ or the $y$ in $\nabla_y$) denote the argument the operators act on. 
For example, $x$ in $\textbf{curl}_x$ means that the curl operator acts on the first variable $x$ in the scalar kernel $\phi (\left | x - y \right |)$.} 

\textcolor{black}{The kernels ${\bf \Phi}^{\text{df}}$ and ${\bf \Phi}^{\text{cf}}$ are called matrix-valued divergence-free and curl-free kernels, respectively, since for any vector $\coeff\in\mathbb{R}^3$, $\nabla_x \cdot \left({\bf \Phi}^{\text{df}}(x,y)\coeff\right)=0$ and $\textbf{curl}_x \left({\bf \Phi}^{\text{cf}}(x,y)\coeff\right)=0$.  This follows by noting that  ${\bf \Phi}^{\text{df}}(x,y)\coeff = \textbf{curl}_x \left[\textbf{curl}_y\left(\phi(|x-y|)\coeff\right)\right]$ and ${\bf \Phi}^{\text{cf}}(x,y)\coeff = \nabla_x \left[\left(\nabla_y\phi(|x-y|)\right)^\intercal\coeff\right]$, i.e., the products are given as the curl of a potential and the gradient of a potential, respectively.  Thus, for a given $y\in\Omega$ and $\coeff\in\mathbb{R}^3$, the product ${\bf \Phi}(x,y)\coeff$ is a vector field in $x$ that can be decomposed as the sum of a divergence-free and curl-free field: ${\bf \Phi}(x,y)\coeff = {\bf \Phi}^{\text{df}}(x,y)\coeff + {\bf \Phi}^{\text{cf}}(x,y)\coeff$, which mimics the HHD.}

\textcolor{black}{We note that the above definitions and notation for ${\bf \Phi}^{\text{df}}$ and ${\bf \Phi}^{\text{cf}}$ in \eqref{eq:div_curl_kernels3d} can be simplified and generalized to bounded domains in $\mathbb{R}^d$, $d=2$ or $3$, as 
\begin{align}
        {\bf \Phi}(x,y) =  \underbrace{(-\Delta{\bf I} + \nabla\nabla^\intercal)\phi(\left|x-y\right|)}_{=: \displaystyle {\bf \Phi}^{\text{df}}(x,y)} - \underbrace{\nabla \nabla^\intercal \phi(\left | x-y \right |)}_{\displaystyle =: {\bf \Phi}^{\text{cf}}(x,y)},\label{eq:div_curl_kernels}
\end{align}
where all differential operators are applied to the $x$ variable and $\nabla\nabla^\intercal$ is the Hessian. This follows by using the result that $\nabla_y \phi(|x-y|) = -\nabla_x\phi(|x-y|)$ and using the standard definition of the curl of a scalar field in $\mathbb{R}^2$.} \textcolor{black}{The scalar radial kernel $\phi:\mathbb{R}^d \times \mathbb{R}^d \to \mathbb{R}$ can take various forms such as Gaussian, multiquadric, or Mat\'ern~\citep{fasshauer2007meshfree}. }

\textcolor{black}{The basic RBF-HHD solver constructs an approximant to the samples of a vector $\mathbf{f}:\mathbb{R}^d \to \mathbb{R}^d$ at the locations $X = \{x_1, x_2, \hdots, x_N\}\subset\Omega$, where $N$ is the number of given data, using the following linear combination of the kernels ${\bf \Phi}^{\text{df}}$ and ${\bf \Phi}^{\text{cf}}$:}
\begin{equation}
\mathbf{s}_{\bf{f}}(x) = \underbrace{\sum_{j=1}^{N} \mathbf{\Phi}^{\rm df}(x,x_j)\coeff_j}_{\displaystyle = \mathbf{s}_{\bf{f}}^{\rm df}(x)} + \underbrace{\sum_{j=1}^{N}\mathbf{\Phi}^{\rm cf}(x,x_j)\coeff_j}_{\displaystyle = \mathbf{s}_{\mathbf{f}}^{\rm cf}(x)} = \sum_{j=1}^{N} \mathbf{\Phi}(x,x_j)\coeff_j, 
\label{eq: decomp}
\end{equation}
where $\coeff_j\in\mathbb{R}^d$ are the expansion coefficients. 
This mimics the HHD of a vector field since by construction $\mathbf{s}_{\mathbf{f}}^{\rm df}$ is divergence-free and $\mathbf{s}_{\mathbf{f}}^{\rm cf}$ is curl-free.  
The expansion coefficients can be determined by forcing $\mathbf{s}_{\mathbf{f}}$ to interpolate the given data, i.e., $\mathbf{s}_{\mathbf{f}}(x_j) = \mathbf{f}_j$, $j=1,\ldots,N$.  
Once this system is solved, the coefficients can be substituted into \eqref{eq: decomp} to obtain a decomposition of $\mathbf{f}$ into divergence-free ($\mathbf{s}_{\mathbf{f}}^{\rm df}$) and curl-free ($\mathbf{s}_{\mathbf{f}}^{\rm cf}$) parts\footnote{In practice, we incorporate boundary conditions into the approximant $\mathbf{s}_{\mathbf{f}}$ to obtain a proper approximate decomposition of $\mathbf{f}$ as described below and further in Appx.~\ref{sect: solvers hhd}.}.

This RBF-HHD solver is fundamentally different from traditional HHD solvers which are usually Poisson-based or use ``combined'' kernels.  For a Poisson-based solver, a vector field ${\bf{s}_f}$ is assumed to have the decomposition
\begin{equation}
    {\bf{s}_f} = -\nabla \varphi + \bf B,
    \label{eq: poisson hhd}
\end{equation}
where $\varphi$ is a scalar potential and $\bf B$ is a divergence-free field. A Poisson equation for $\varphi$ is formulated by taking divergence on both sides of \eqref{eq: poisson hhd}:
\begin{equation}
    \nabla \cdot {\bf{s}_f} = - \nabla^2 \varphi + \nabla \cdot \bf B = - \nabla^2 \varphi,
    \label{eq: poisson-based ppe}
\end{equation}
subject to some boundary conditions.
After solving \eqref{eq: poisson-based ppe} using the given data $\nabla \cdot \left . {\bf{s}_f} \right |_X = \nabla \cdot \left . {\bf f} \right |_X$, we can obtain the scalar $\varphi$ and hence the curl-free part $\nabla \varphi$.
The divergence-free field is then given as $\mathbf{B} = {\bf{s}_f} - \nabla \varphi$. Such a Poisson-based HHD solver may suffer from imposing incorrect boundary conditions for the scalar $\varphi$ \citep{fuselier2017radial}.  Additionally, if $\mathbf{f}$ is noisy then this noise can be amplified by computing $\nabla\cdot\mathbf{f}$.

For a ``combined'' kernel method (see~\citet{schrader2011high,wendland2009divergence} for example), an approximant is used that combines the divergence-free kernel ${\bf \Phi}^\text{df}$ from \eqref{eq:div_curl_kernels} with a scalar kernel $\psi (\left | x-y \right |)$ as
\begin{equation*}
    \tilde{\bf \Phi} = \begin{bmatrix}
        {\bf \Phi}^\text{df} & {\bf 0} \\ 
        {\bf 0} & \psi
    \end{bmatrix},
\end{equation*}
which models the decomposed vector field using a divergence-free kernel ${\bf \Phi}^\text{df}$ and a scalar curl-free potential $\psi$.  This type of solver usually has a higher computational cost \textcolor{black}{than that} of the RBF-HHD solver \citep{fuselier2017radial}.

We can construct a divergence-free RBF-HHD solver for velocity field reconstruction for incompressible flows and a curl-free RBF-HHD solver for pressure field reconstruction based on the above decomposition.
When solving for  ${\coeff}_j$, 
the divergence-free solver utilizes $\mathbf{s}_{\mathbf{f}}^{\rm df}$ 
from \eqref{eq: decomp} and its boundary conditions are prescribed as the outward normal components of the divergence-free part of $\mathbf{f}$ at the boundaries.
On the other hand, the curl-free solver employs $\mathbf{s}_{\mathbf{f}}^{\rm cf}$ and its boundary conditions are imposed to be the tangential components of the curl-free part of $\mathbf{f}$ at the boundaries.
After the decomposition, the divergence-free RBF-HHD solver removes the curl-free bias in the velocity field for an incompressible flow.
\textcolor{black}{The curl-free RBF-HHD solver filters out the divergence-free component in the pressure gradients and calculates a potential for the pressure gradients (i.e., the pressure field).
The potential is recovered from the sum of the terms $-(\nabla\phi(\left | x-x_j \right|))^\intercal \coeff_j$.}

The RBF-HHD solver is versatile. 
First, the solver can suppress the impact of random noise in the reconstruction by framing the RBF-HHD into a regression-based solver. 
Second, the solver can remove bias in the input vector data, such as the curl-free part in corrupted velocity fields and the divergence-free part in contaminated pressure gradient fields.
Third, we can directly recover the potential functions of a vector field (e.g., reconstruct the pressure field from the pressure gradients). Lastly, the solver is able to approximate scattered data in a complex domain.

\section{Validation}
\label{sec: validation}
In this section, we use synthetic PIV and PTV data generated by adding artificial errors to the ground truth (from exact solutions or high-fidelity simulations) to test the performance of several pressure solvers and validate the arguments and predictions introduced in Sect.~\ref{Sec: Insights from the error estimation}. 
We also demonstrate the performance of the RBF-HHD solvers presented in Sect.~\ref{sec: RBF-HHD solver}.

\subsection{Tests on PGI solvers (RBF-HHD and RPR-ODI)}
\label{Sec: HHD vs ODI}
\begin{figure}[!htb]
	\centering
	\includegraphics[width=1\columnwidth]{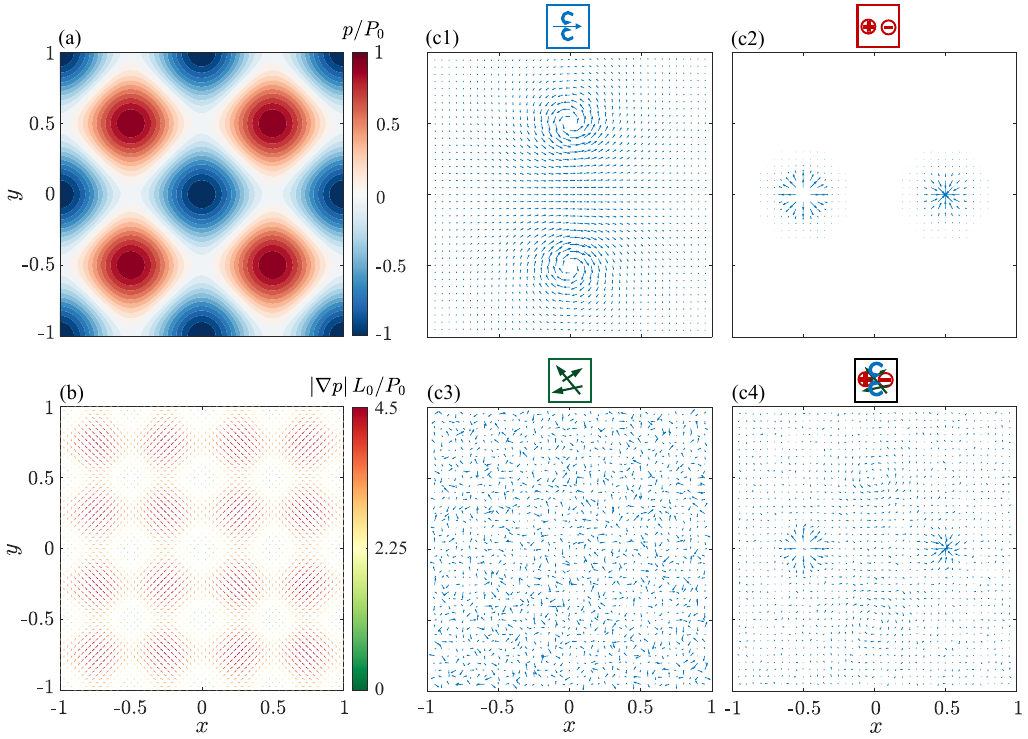}
	\caption{The pressure and its gradients of the Taylor-Green vortex, and the added error in the PGI solver test. (a)~The true value of the normalized pressure field $p/P_0$; (b) the true value of the pressure gradients, colored by $\left | \nabla p \right | L_0/P_0$. (c1)~--~(c4): errors $\epsilon_{\nabla p}$ superposed in the input pressure gradient: (c1) the divergence-free error (case~1), (c2) the curl-free error (case~2), (c3) the random error (case~3), and (c4) the compound error (case~4). (b) and (c1)~--~(c4) are down-sampled to half of the actual resolution for visualization purposes.}
\label{fig: tgvinput}
\end{figure}

We used synthetic pressure gradient data to test the properties of two PGI solvers, one is the RBF-HHD solver presented in Sect.~\ref{sec: RBF-HHD solver}, and the other is the RPR-ODI, which is the state-of-the-art of the ODI family. 
The synthetic data were based on a 2D Taylor-Green Vortex (TGV), in the domain $(x,y) \in [-1,1] \times [-1,1]$.
The data points are on a uniform Cartesian mesh, with 81 nodal points in each direction.
The grid spacing in the $x$ and $y$ directions is the same.
The pressure ground truth of the TGV is given by:
$$p = -\frac{1}{4}\left(\cos(2n\pi x) + \cos(2n\pi y)\right),$$
where $n=1$ is the wave number, {\color{black}{the fluid density is set to be unity}}, and the corresponding gradient ground truth $\nabla p = (\partial_x p, \partial_y p)$ can be explicitly evaluated (see Fig.~\ref{fig: tgvinput}(a) and (b) for illustration).

Four sets of artificial errors were added to the ground truth of the pressure gradient to generate the synthetic data. 
First, a divergence-free error was constructed by \textcolor{black}{superposing two vortex-shaped errors}, $\bm \epsilon_1$ and $\bm \epsilon_2$, centered at $\bm{r}_1$ and $\bm{r}_2$ spinning in opposite directions: 
\begin{equation*}
    \bm \epsilon^{\text{df}}_{\nabla p} = \bm \epsilon_1 + \bm \epsilon_2,
\end{equation*}
where the velocity components of $\bm{\epsilon}_{1}$ and $\bm{\epsilon}_{2}$ in polar coordinates $(r,\theta)$ are $\epsilon_{r}(r_{1,2}) = 0$, $\epsilon_{\theta}(r_{1,2}) = \pm \frac{C_1}{r_{1,2}}(1-\exp({-r_{1,2}^2}))$, and $r_{1,2} = C_2\sqrt{x^2 + (y \pm y_0)^2}$. 
This error looks like the cross-section of a vortex ring with a `jet' shooting from left to right (see Fig.~\ref{fig: tgvinput}(c1) and the icon for symbolic representation). 
Second, a curl-free error in the pressure gradient field $\bm \epsilon^{\text{cf}}_{{\nabla p}}$ was constructed by taking the gradient of a scalar field
\begin{equation*}
    \bm \epsilon^{\text{cf}}_{{\nabla p}} = \nabla (Q_3 + Q_4),
\end{equation*}
where $Q_{3,4} = \pm C_3\exp(-r_{3,4}^2),$ and $r_{3,4}=C_4\sqrt{(x \pm x_0)^2 + (y)^2}$. 
This error looks like a sink/source (Fig.~\ref{fig: tgvinput}(c2) and the icon).
Both $\bm \epsilon^{\text{df}}_{\nabla p}$ and $\bm \epsilon^{\text{cf}}_{{\nabla p}}$ \textcolor{black}{were} mean zero by symmetry. 
Third, we constructed point-wise random error 
\begin{equation*}
    \bm \epsilon^{\text{rn}}_{{\nabla p}} \sim \mathcal{N} \left ( \begin{bmatrix}
0 \\ 
0
\end{bmatrix},
\begin{bmatrix}
\sigma^2  & 0 \\ 
0 & \sigma^2 
\end{bmatrix} \right ),
\end{equation*}
which \textcolor{black}{was} an independent and identically distributed Gaussian noise with a variance of $\sigma^2$ and zero mean in the $x$ and $y$ directions (see Fig.~\ref{fig: tgvinput}(c3) and the icon).
Last, the compound error was a super-position of the above three errors:
\begin{equation*}
    \bm \epsilon^{\text{cmp}}_{{\nabla p}} =\bm \epsilon^{\text{df}}_{\nabla p} + \bm \epsilon^{\text{cf}}_{{\nabla p}} +  \bm \epsilon^{\text{rn}}_{{\nabla p}},
\end{equation*}
and is visualized in Fig.~\ref{fig: tgvinput}(c4) with a corresponding icon.

In this work, we define a space-averaged $L^2$-norm of a quantity $[\cdot]$ as
\begin{equation}
    \left \| [\cdot]  \right \|_{L^2(\Omega)} = \sqrt{\frac{\int [\cdot]^2 d \Omega}{|\Omega|}},
    \label{eq: l2 norm}
\end{equation}
where $|\Omega|$ is the area or volume of the domain.
The parameters of the first three errors in the input data were adjusted so that their $L^2$ norm ($\| \bm \epsilon_{\nabla p} \|_{L^2(\Omega)}$) were close to 40\% of the norm of the pressure gradient, i.e., $\| \bm \epsilon^{\text{df}}_{\nabla p} \|_{L^2(\Omega)} \approx  \| \bm \epsilon^{\text{cf}}_{\nabla p} \|_{L^2(\Omega)} \approx  \| \bm \epsilon^{\text{rn}}_{\nabla p} \|_{L^2(\Omega)} \approx  0.4\| \nabla p \|_{L^2(\Omega)}$, leading to $\| \bm \epsilon^{\text{cmp}}_{\nabla p} \|_{L^2(\Omega)} \approx  0.7\| \nabla p \|_{L^2(\Omega)}$. 
In addition, the spatial length scales of the errors ($\bm \epsilon^{\text{df}}_{\nabla p}$ and $\bm \epsilon^{\text{cf}}_{\nabla p}$) were kept similar.
The values of the parameters used for generating the errors are listed in Table~\ref{tab: parameters for errors}.
We added these artificial errors to contaminate the ground truth of $\nabla p$, and the resulting four sets of synthetic data are
case~1: $ \bm{\tilde g}_1 = \nabla p + \bm \epsilon^{\text{df}}_{\nabla p} $; case~2: $ \bm{\tilde g}_2 = \nabla p + \bm \epsilon^{\text{cf}}_{\nabla p} $; case~3: $ \bm{\tilde g}_3 = \nabla p + \bm \epsilon^{\text{rn}}_{\nabla p} $; and case~4: $ \bm{\tilde g}_4 = \nabla p + \bm \epsilon^{\text{cmp}}_{\nabla p}$. 
We used characteristic length scale $L_0=1$ and characteristic pressure $P_0 = 0.5$ to normalize the length of the domain, the pressure, the pressure gradient, as well as the pressure reconstruction errors.

\begin{table}[h]
\caption{Parameters of the artificial error used in the validation based on the Taylor-Green vortex.}
\label{tab: parameters for errors}
\centering
\begin{tabular}{llllllllll}
\hline
~ & $x_0$ & $y_0$ & $C_1$ & $C_2$ & $C_3$ & $C_4$ & $\sigma$ & $\| \bm \epsilon_{\nabla p} \|_{L^2(\Omega)}$ & $\| \nabla p \|_{L^2(\Omega)}$ \\ \hline
case 1, div-free error & 0 & 0.5 & 2.5 & 10 & -- & -- & -- & 0.6329 & 1.5806 \\
case 2, curl-free error & 0.5 & 0 & -- & -- & 3.6 & 10 & -- & 0.6381 & 1.5806 \\
case 3, random error& -- & -- & -- & -- & -- & -- & 0.44 & 0.6298 & 1.5806\\  
case 4, compound error & 0.5 & 0.5 & 2.5 & 10 & 3.6 & 10 & 0.44 & 1.0961 & 1.5806 \\ \hline
\end{tabular}
\end{table}


Taking these four contaminated pressure gradient fields as input, we reconstructed the pressure field using the RBF-HHD and RPR-ODI solvers. 
For the RBF-HHD solver, the over-sampling ratio for least squares regression was 20, where the oversampling ratio refers to the ratio of the number of given data points to the number of kernels.
For the RPR-ODI solver, \textcolor{black}{the iterations were terminated either when the number of iterations reached 1,000 or when the residual was below $1\times 10^{-10}$}; the rotation angle increment was $\Delta \alpha = 0.2^{\circ}$, and the distance between each ray normalized by grid spacing was $\Delta d^*=0.2$. 
The above RPR-ODI parameters are recommended by \citet{liu2016instantaneous}.

\begin{figure}[!htb]
	\centering
	\includegraphics[width=1\columnwidth]{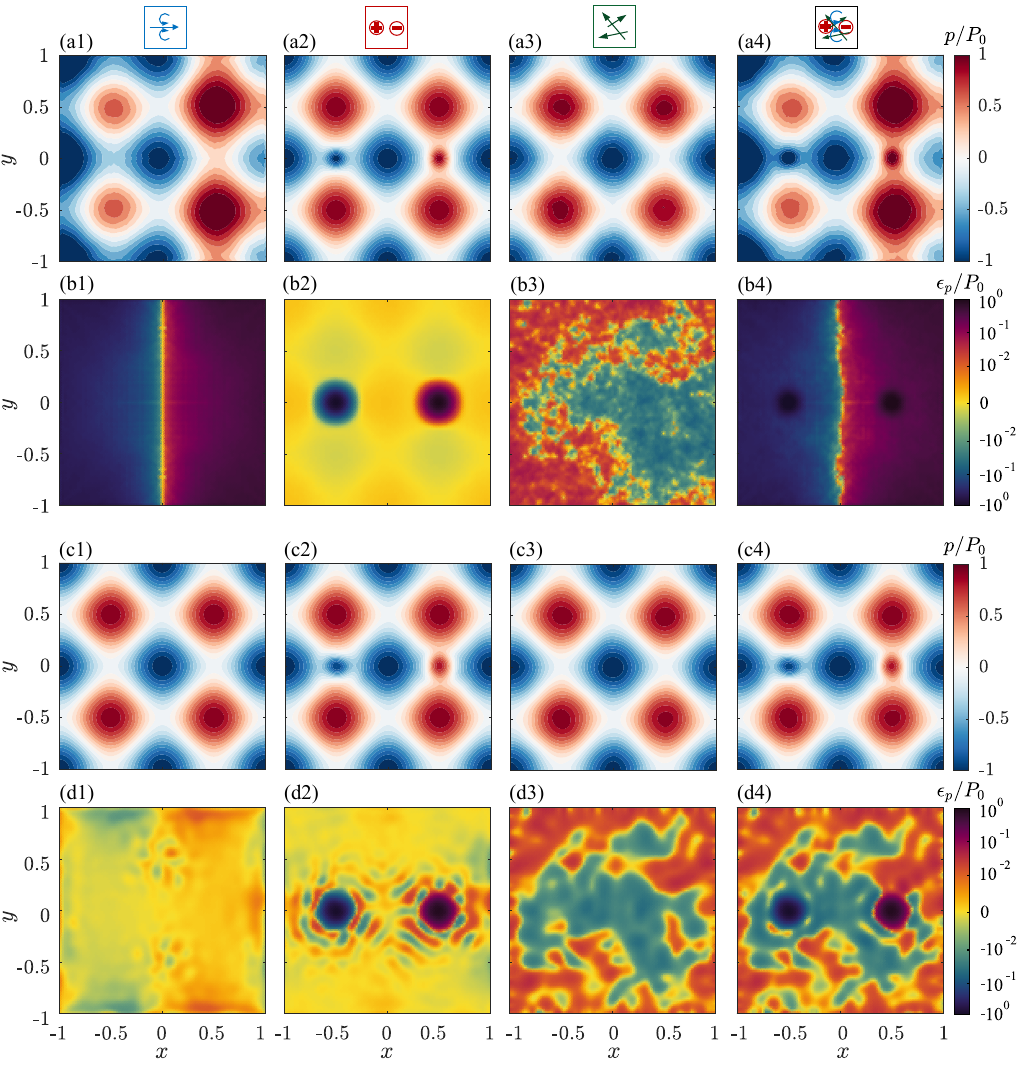}
	\caption{Reconstructed pressure fields and their errors using the RPR-ODI and RBF-HHD solvers. From left column (a1~--~d1) to right (a4~--~d4): the reconstruction based on the input data that are contaminated by the divergence-free (case~1), curl-free (case~2), random (case~3), and compound error (case~4), respectively. From the top two rows (a1~--~b4) to bottom (c1~--~d4): the reconstruction and its errors using the RPR-ODI and RBF-HHD solver, respectively. The pressure and its error are normalized by the characterized pressure $P_0$.}
\label{fig: tgvoutput}
\end{figure}

\begin{figure}[!htb]
	\centering
	\includegraphics[width=0.5\columnwidth]{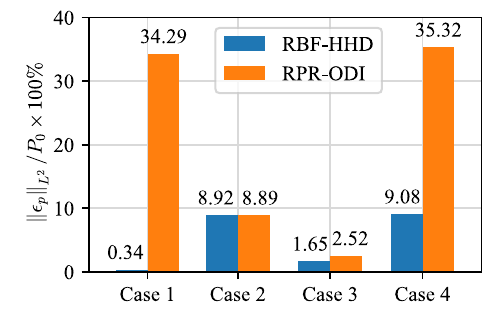}
	\caption{Normalized relative errors in the reconstructed pressure fields using the RPR-ODI and RBF-HHD solvers. The reconstructed pressure fields are based on the input pressure gradients that are contaminated by divergence-free error (case~1), curl-free error (case~2), random error (case~3), and compound error (case~4).}
\label{fig: errorBar}
\end{figure}

The reconstruction results are presented in Figs.~\ref{fig: tgvoutput} and \ref{fig: errorBar}.
Figure~\ref{fig: tgvoutput}(a1)~--~(d1) illustrates the reconstruction results where the divergence-free error was used to contaminate the data (i.e., $\bm{\tilde g}_1 = \nabla p + \bm \epsilon^{\text{df}}_{\nabla p}$). 
As shown in (a1), apparent deviations between the reconstructed pressure field and the ground truth (see Fig.~\ref{fig: tgvinput}(a)) emerge when using the RPR-ODI for reconstruction. 
The error in the pressure reconstruction appears similar to a plane tilting up from left to right.
This phenomenon arises from the divergence-free error in the pressure gradient resembling a `jet' shooting from left to right across the domain, and the RPR-ODI faithfully integrated this bias, resulting in a tilted pressure field.
On the contrary, the reconstruction using the RBF-HHD solver (see (c1)) closely matches the ground truth (Fig.~\ref{fig: tgvinput}(a)), and almost no errors appear in Fig.~\ref{fig: tgvoutput}(d1).
Figure~\ref{fig: errorBar} case~1 shows the quantitative results of the reconstruction errors:
the RPR-ODI exhibits a 34.29\% error compared with 0.34\% when using the RBF-HHD.
This indicates that the RBF-HHD effectively mitigates the divergence-free errors, leading to an accurate pressure field reconstruction.

Figure~\ref{fig: tgvoutput}(a2)~--~(d2) depicts the reconstruction using the curl-free error to corrupt the pressure gradient input (i.e., $\bm{\tilde g}_2 = \nabla p + \bm \epsilon^{\text{cf}}_{\nabla p}$).
An erroneous dent and bump in the reconstructed pressure field emerge near $(x,y)=(\pm0.5,0)$, see (a2) and (b2).
Similar dent and bump of errors are observed when using the RBF-HHD solver (see (c2) and (d2)).
The curl-free error in the pressure gradient is characterized by the pressure gradients pointing outward and inward, resembling a source and sink, respectively. 
The RPR-ODI and RBF-HHD solvers integrate these erroneous gradients faithfully, leading to the nonphysical pressure dent (low pressure) and bump (high pressure) in (a2) and (b2).
In Fig.~\ref{fig: errorBar} case~2, we see that both solvers yield almost the same level of reconstruction error, which is about 8.9\%.
This parity in performance shows that both solvers cannot detect or remove curl-free errors.

Figure~\ref{fig: tgvoutput}(a3)~--~(d3) illustrates the reconstruction using data with added random errors (i.e., $\bm{\tilde g}_3 = \nabla p + \bm \epsilon^{\text{rn}}_{\nabla p}$).
In (a3), no discernible differences are observed between the pressure reconstruction and its ground truth when using the RPR-ODI.
Its reconstruction errors are low in almost the entire domain (see (b3)).
We can see similar results when using the RBF-HHD solver.
Although both solvers do well in filtering out random errors, the RBF-HHD solver performs slightly better than the RPR-ODI.
This is evidenced by the fact that the RBF-HHD solver \textcolor{black}{yielded} a reconstruction error that was about one-third lower than that of the RPR-ODI solver (see Fig.~\ref{fig: errorBar} case~3). 
This one-third difference in reconstruction errors may be attributed to the composition of random error and the nature of the RBF-HHD solver.
The random error may consist of three equal portions of the divergence-free, curl-free, and harmonic components; and the RBF-HHD solver can completely remove the divergence-free component while the RPR-ODI cannot.

Figure~\ref{fig: tgvoutput}(a4)~--~(d4) displays the pressure reconstruction and the corresponding error using the pressure gradient data contaminated by the compound errors (i.e., $\bm{\tilde{g}}_4 = \nabla p + \bm \epsilon^{\text{df}}_{\nabla p} + \bm \epsilon^{\text{cf}}_{\nabla p} + \bm \epsilon^{\text{rn}}_{\nabla p}$).
Both RPR-ODI and RBF-HHD solvers give smooth pressure reconstruction, implying that both solvers enjoy robustness against random noise.
In (a4), the divergence-free and curl-free errors penetrate from the pressure gradients to pressure fields when using the RPR-ODI solver, while only the curl-free error penetrates the pressure fields when using the RBF-HHD solver.
As shown in Fig.~\ref{fig: errorBar} case~4, the RBF-HHD solver has a lower overall reconstruction error than that of the RPR-ODI (9.08\% vs. 35.32\%).

From the above test, we find that the RBF-HHD solver can almost completely remove the divergence-free bias in the input pressure gradient data.
Besides, the RBF-HHD solver is robust to random noise.
The test validates the arguments in Sect.~\ref{Sec: HHD-based regularization} and demonstrates superior performance of the RBF-HHD solver discussed in Sect.~\ref{Sec: HHD as a continuous limit of ODI}.

\subsection{Validation for the scaling laws (PPE and PGI)}
\label{sec: scaling law validation}

In this section, we validate the scaling laws discussed in Sect.~\ref{Sec: Error scaling for PPE and PGI}, concerning the number of nodal points $N$ in one direction, the length scale of the domain $L$, and the standard deviation $\sigma$ of a  zero-mean Gaussian noise.

We consider a `realistic' Taylor vortex in water to validate the scaling laws.
Assuming the pressure vanishes in the far field, the pressure field of the vortex is given by \citep{panton2006incompressible,charonko2010assessment}:
\begin{equation}
    p(r,t) = \frac{\rho H^2}{64 \pi^2 \nu t^3} \text{exp} (-\frac{r^2}{2 \nu t}), \label{eq: tv pressure}
\end{equation}
where $H=7.5 \times 10^{-5}~\text{m}^2$ represents angular momentum of the vortex, $\nu=1 \times 10^{-6}~\text{m}^2~\text{s}^{-1}$ is the kinematic viscosity of water, $\rho=1 \times 10^3~\text{kg}~\text{m}^{-3}$ is the density of water, $r$ is the distance between a data point in the vortex to the vortex center, and the time $t = 0.5~\text{s}$.
These parameters were chosen following \citet{mcclure2017instantaneous}.
The characteristic length $L_0$, velocity $U_0$, and pressure $p_0$ are defined as $L_0=\sqrt{2 \nu t}$, $U_0=H/(2 \pi L_0 t)$, and $P_0=\rho U_0^2$, respectively.
The pressure $p$, pressure gradients $\nabla p$, and pressure Laplacian $\nabla^2 p$ are non-dimensionalized as $p^* = p/P_0$, $\nabla p^* = \nabla p L_0/P_0$, and $\nabla^2 p^* = \nabla^2 p L_0^2/P_0$, respectively. 
The sizes of the domain is non-dimensionalized as $x^* = x/L_0$ and $y^* = y/L_0$.

To generate the synthetic data, we imposed artificial point-wise noise on the pressure gradient ground truth and varied $N$, $L$, and $\sigma$ in the synthetic data.
The ground truth of the pressure and its gradients were analytically evaluated based on \eqref{eq: tv pressure}.
For the $N$ scaling law tests, we varied $N$ in the data, from 20 to 240.
The variation in $N$ essentially changes the spatial resolution of the data.
The non-dimensional length of the domain was set to $L^* = L/L_0 = 8$.
The artificial noise imposed on the data was a zero-mean Gaussian noise, and its standard deviation was adjusted to make the power of the noise almost the same and approximately 10\% of the pressure gradients, i.e.,  $\| \bm \epsilon_{\nabla p} \|_{L^2(\Omega)} \approx \| \bm \epsilon_{\nabla^2 p} \|_{L^2(\Omega)} \approx 0.1 \| \nabla p \|_{L^2(\Omega)}$.
We also used noise-free data for the grid convergence test to highlight the difference between our error scaling with respect to $N$ and the solvers' intrinsic order of accuracy.

For the $L$ scaling law test, we varied $L^*$ from 6 to 48, and $N$ was set to 200.
The artificial noise was also zero-mean Gaussian noise, and its standard deviation was adjusted to make the normalized power of the noise about 0.01, i.e., $\| \bm \epsilon_{\nabla p} \|_{L^2(\Omega)} \approx \| \bm \epsilon_{\nabla^2 p} \|_{L^2(\Omega)} \approx 0.01$.
For the $\sigma$ scaling law tests, we set $N = 100$ and $L^* = 8$ in the data.
The standard deviation of the zero-mean Gaussian noise varied from $10^{-7}$ to 1.

Three categories of pressure solvers were used in this validation.
The first category includes second- and fourth-order finite difference Poisson solvers with Dirichlet or Neumann boundary conditions imposed.
The second category is the RPR-ODI solver.
In the RPR-ODI solver, the number of iterations was set to 20 to \textcolor{black}{accelerate computation}. 
The rotation angle increment was $\Delta \alpha = 0.5^{\circ}$; and the distance between rays normalized by the grid spacing was $\Delta d^*=0.5$, which \textcolor{black}{are considered as medium-accuracy RPR-ODI reconstruction} \citep{liu2016instantaneous}.
The third is two HHD solvers. 
They are the RBF-HHD~\citep{fuselier2017radial} and a PPE-based HHD (dubbed as PPE-HHD)~\citep{mcclure2017instantaneous}. 
Details about the solvers can be found in Appx.~\ref{sect: solvers} for convenience or the corresponding original papers.
\textcolor{black}{In each scaling law test, five independent tests were repeated for each solver under every parameter configuration.}

\begin{figure}[!htb]
	\centering
    \includegraphics[width = 0.9\textwidth]{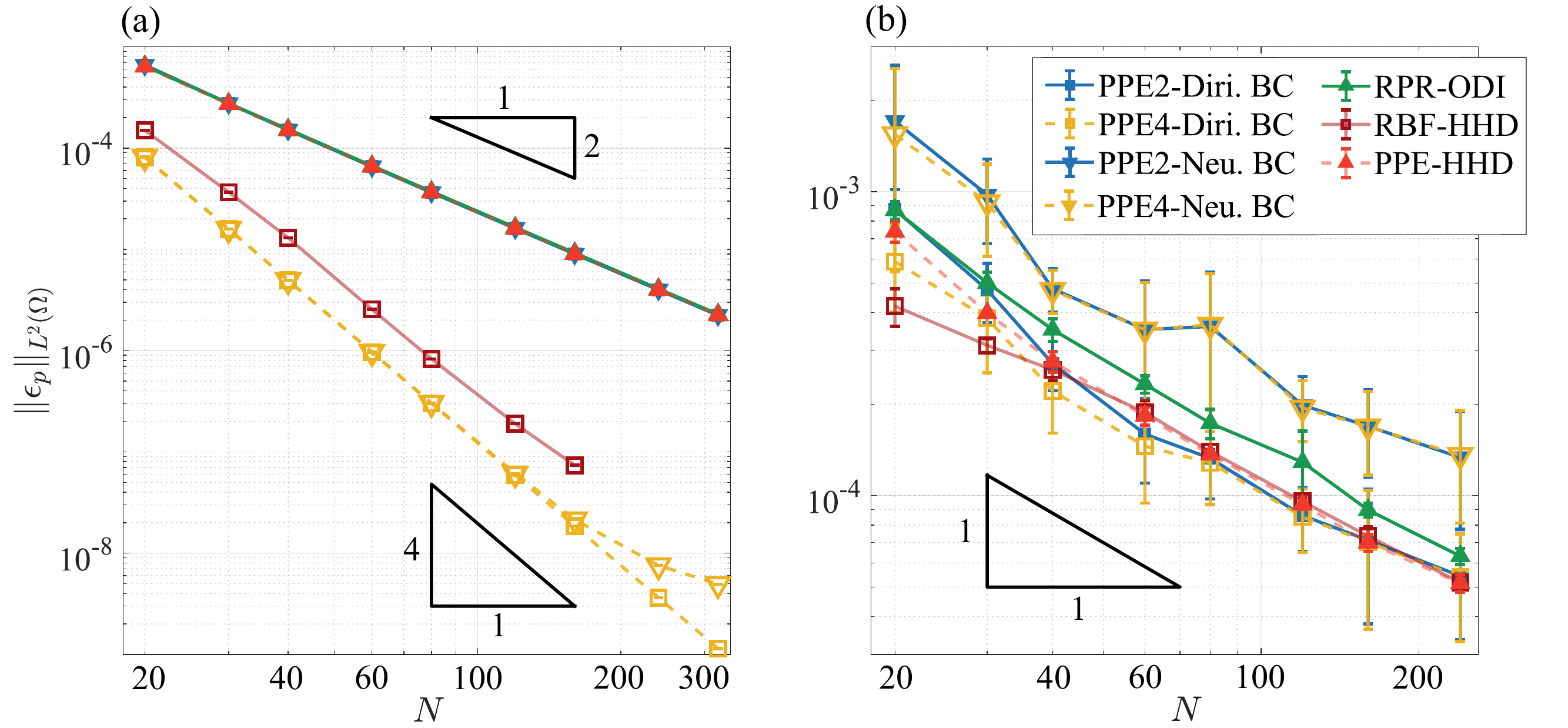}
	\caption{Non-dimensional errors in the reconstructed pressure field vs. the number of nodal points in one direction ($N$) without (a) and with (b) the Gaussian noise added to the data. The PPE2-Diri.~BC, PPE4-Diri.~BC, PPE2-Neu.~BC, and PPE4-Neu.~BC represent the second- and fourth-order finite difference pressure Poisson solvers, with Dirichlet (Diri.~BC) or Neumann (Neu.~BC) boundary conditions, respectively. \textcolor{black}{Each error represents the sample statistics for independently repeated tests. The triangle or square symbols indicate the mean error. The upper and lower whiskers indicate the standard deviation of the error. The same fashion of the error bars is also applied to Figs.~\ref{fig: l_scaling} and \ref{fig: s_scaling}.}
 }
\label{fig: n_scaling}
\end{figure}

\begin{figure}[!htb]
	\centering
	\includegraphics[scale=0.36]{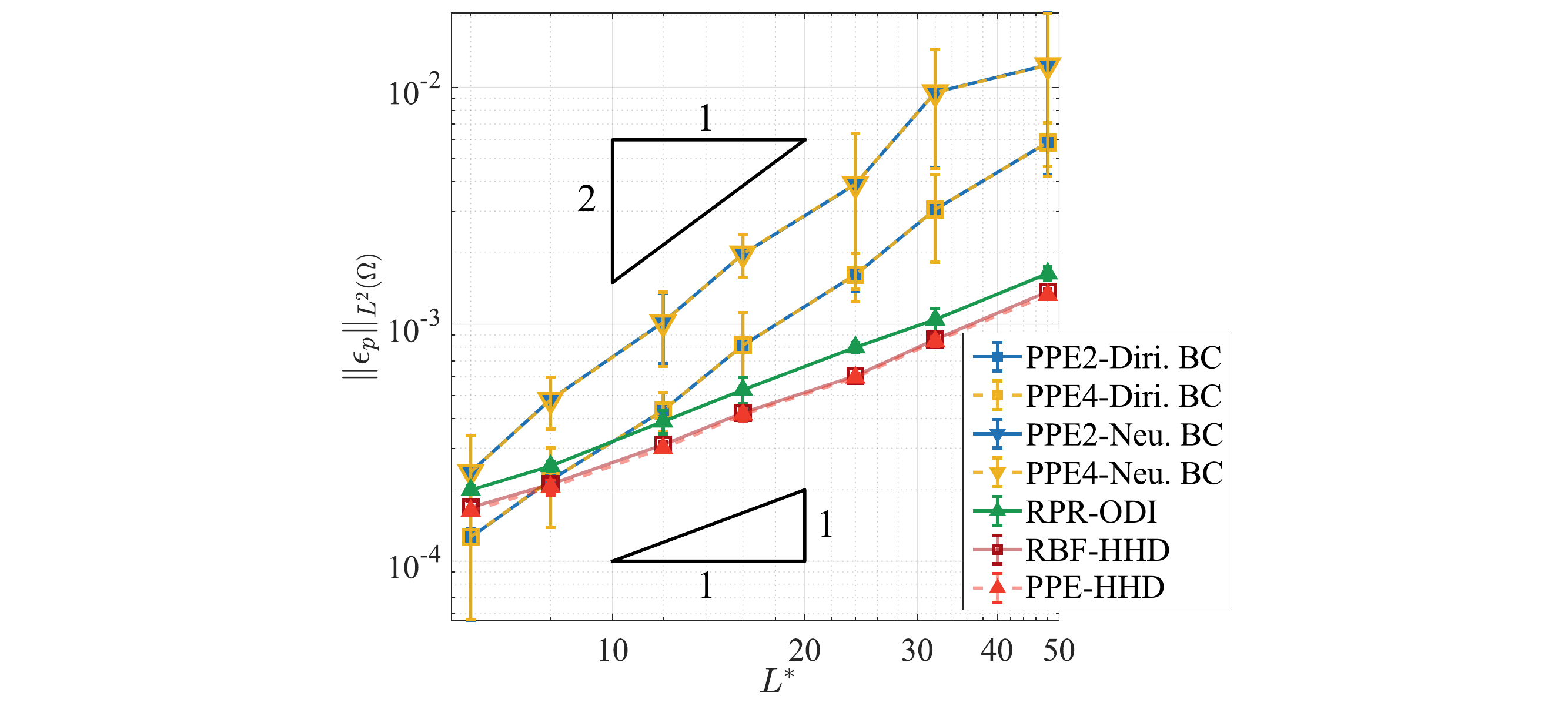}
	\caption{Non-dimensional errors in the reconstructed pressure field vs. non-dimensional length scale of the \textcolor{black}{domain ($L^*$)}. 
 }
\label{fig: l_scaling}
\end{figure}

\begin{figure}[!htb]
	\centering
	\includegraphics[scale=0.36]{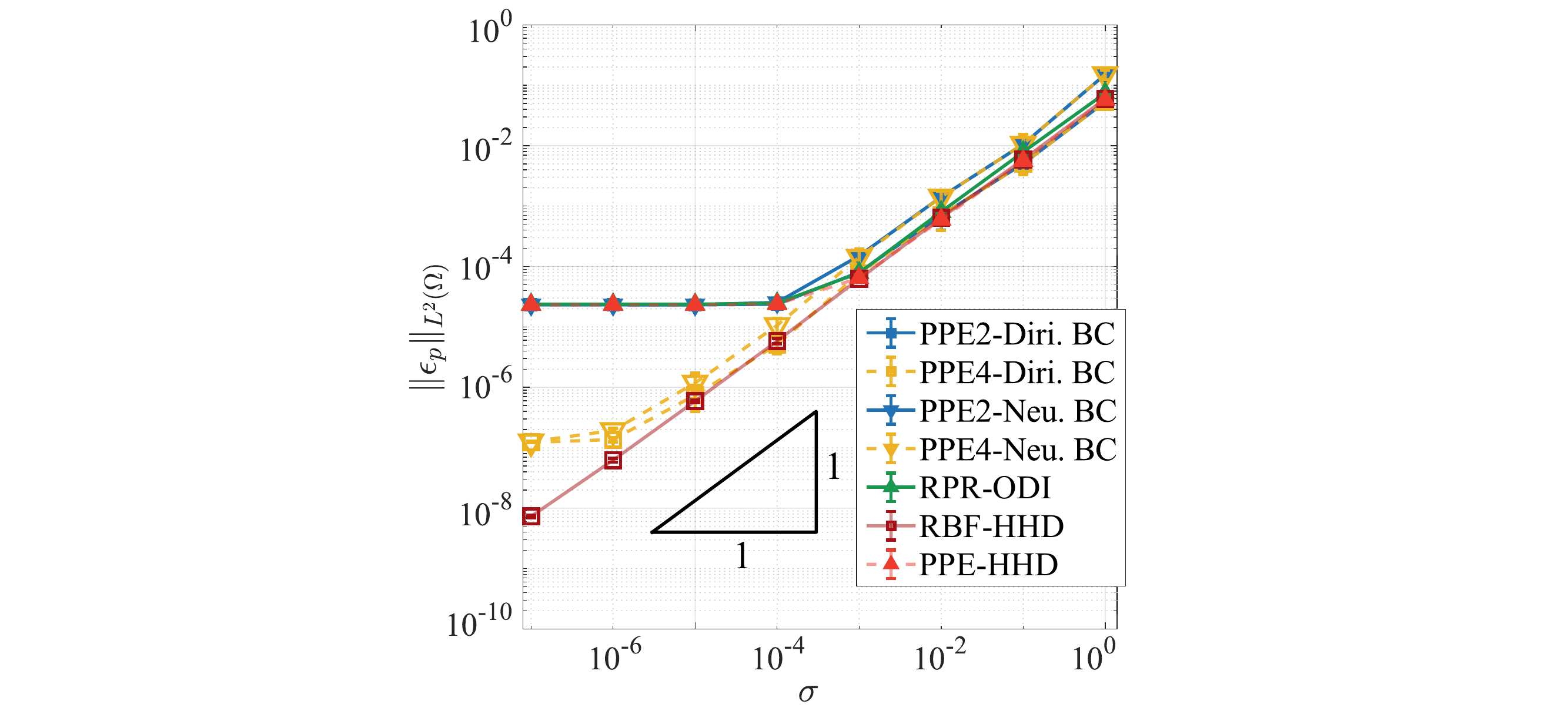}
	\caption{Non-dimensional errors in the reconstructed pressure field vs. the standard deviation of the zero-mean Gaussian noise ($\sigma$). 
 }
\label{fig: s_scaling}
\end{figure}

The validation results for the scaling laws are presented in Figs.~\ref{fig: n_scaling}~--~\ref{fig: s_scaling}.
As shown in the grid convergence tests (see Fig.~\ref{fig: n_scaling}(a)), when the noise was not imposed on the data, numerical solvers recovered their intended orders of accuracy.
For example, the second- and fourth-order finite difference based solvers show trends with slopes of $-2$ and $-4$, respectively, regardless of the type of boundary conditions.
Both the trends for the RPR-ODI and PPE-HHD have a slope of $-2$ convergence due to the embedded second-order numerical differentiation \citep{liu2020error,mcclure2017instantaneous}. 
The RBF-HHD solver shows a trend with a slope close to $-4$, suggesting that it is a high-order solver with about fourth-order accuracy.

Figure~\ref{fig: n_scaling}(b) illustrates the scaling of $\| \epsilon_{p}\|_{L^2(\Omega)}$ with respect to the resolution of the corrupted data.
As shown in this sub-figure, the trends of the tests all experience a slope of $-1$ for all solvers when the noise is presented in the data ($f$ or $\bm{g}$).
This prediction is in good agreement with \eqref{eq: scaling law PPE simple main text} and \eqref{eq: scaling law dpgi simple main text}: the PGI or PPE solvers all follow the $N^{-1}$ scaling law.

The scaling of the error in the reconstructed pressure with respect to the length scale of the domain is depicted in Fig.~\ref{fig: l_scaling}. 
The slopes are about $+1$ for the PGI solvers and $+2$ for the PPE solvers.
These observations agree with what we predicted by \eqref{eq: scaling law PPE simple main text} and \eqref{eq: scaling law dpgi simple main text}, which are rooted in the fact that the solver either integrates once or twice to obtain the pressure from $\nabla p$ or $\nabla^2 p$, respectively.

Figure~\ref{fig: s_scaling} presents the scaling of the error in the reconstructed pressure relative to the standard deviation $\sigma$.
When $\sigma$ is sufficiently large (i.e., $\sigma > 10^{-3}$), the slope of the trend with regard to $N$ is nearly +1 regardless of the choice of pressure solvers.
Equations \eqref{eq: scaling law PPE simple main text} and \eqref{eq: scaling law dpgi simple main text} confirm this, indicating that all PGI or PPE solvers follow a +1 scaling law with respect to $\sigma$.
In Fig.~\ref{fig: s_scaling}, we also observe that when the noise in the input data is sufficiently small (e.g., $\sigma < 10^{-3}$ for the second-order PPE solvers), the slopes of these curves change from +1 to zero. 
This change in the error could be attributed to the diminishing influence of the small random noise in the data, which no longer outweighs the truncation errors embedded within the numerical solvers. 
Consequently, the reconstruction error becomes primarily dependent on the spatial resolution and the order of accuracy of the numerical scheme, rather than any explicitly added error.
This explanation is further supported by the observation that, in the high-order solvers such as the RBF-HHD and fourth-order finite difference solvers, the error plateau onset is delayed as $\sigma$ decreases compared to the low-order ones.

From the above tests, we find that the general nature of the error aligns with the predictions made in Sect.~\ref{Sec: Error scaling for PPE and PGI}, confirming the accuracy of the scaling laws derived there. 
Moreover, as shown in Figs.~\ref{fig: n_scaling}~--~\ref{fig: s_scaling}, the green lines that represent the errors from the RPR-ODI solvers almost always lie above those of the RBF-HHD solver, i.e., the RBF-HHD solver is consistently more accurate than the RPR-ODI.
This observation implies that the RBF-HHD solver outperforms the RPR-ODI solver, which is additional support for the argument in Sect.~\ref{Sec: HHD as a continuous limit of ODI}, and consistent with the conclusion in Sect.~\ref{Sec: HHD vs ODI} based on careful benchmarking with a different flow scenario.

\subsection{Demonstration of the RBF-HHD on Complex Flow}
\label{sec: cylinder flow test}
A two-dimensional, laminar, unsteady flow around a circular cylinder for a Reynolds number $Re=100$ is used to showcase our solvers and demonstrate some of the arguments in Sect.~\ref{Sec: Insights from the error estimation}.

We first used a high-fidelity numerical simulation to generate the ground truth of the flow field. 
The numerical simulation was conducted in OpenFOAM using the \textit{icoFoam} solver~\citep{weller1998tensorial}.
The free stream velocity of the flow was $U_\infty = 1$, the kinematic viscosity was $\nu = 0.01$, and the density was $\rho = 1$.
The numerical simulation domain spanned $(x/D,y/D) \in [-8,25] \times [-8,8]$, in which the center of the cylinder with diameter $D = 1$ placed at $(x/D,y/D) = (0,0)$.
The boundary conditions included a velocity inlet at $x/D=-8$, a pressure outlet $p = 0$ at $x/D=25$, symmetry planes at $y/D=\pm 8$, and a non-slip wall at the cylinder. 
We used a structured mesh with about 0.75 million cells to discretize the domain, with mesh refinement near the cylinder and wake region.

The validation was performed within a rectangular reconstruction domain that was cropped from the simulation domain.
The reconstruction domain enclosed the stagnation region in front of the cylinder and the oscillating wake behind it.
This domain spanned $(x/D,y/D) \in [-1,3] \times [-1,1]$.
We used $U_\infty$, $D$, and $P_{\infty} = \frac{1}{2} \rho U^2_{\infty}$ as the characteristic scales to normalize the velocity, pressure, and pressure gradients, respectively, as well as the corresponding reconstruction errors.

To emulate realistic PTV experiments, synthetic Lagrangian data were generated by imposing Gaussian noise on the true value of the velocity sampled at pseudo-particles in the domain.
About $1.1\times10^4$ pathlines originated at random and unique locations were generated by integrating the instantaneous velocity.
The instantaneous velocity and pressure at the pseudo-particle locations along the pathlines were interpolated using the data from the simulation, serving as the ground truth for validation. 
Last, we imposed zero-mean Gaussian noise on the particle velocity. 
This artificially corrupted velocity field was used as the synthetic PTV data.
The standard deviation of the artificial noise was equal to $ 1\% $ or $ 5\% $ of the magnitude of the local velocity.

To reconstruct the pressure fields from the velocity fields, we first evaluate the pressure gradients from the synthetic Lagrangian data, using the material acceleration and viscosity terms.
The material acceleration is calculated by approximating the velocity change when following each particle pathline \citep{van2013piv}, using three equal-spacing consecutive frames and the second-order central finite difference scheme:
\begin{equation*}
    \frac{D {\bm{u}}({\bm {x}}(t),t)}{D t} \approx \frac{{\bm{u}}(t+ \Delta t ) - {\bm{u}}(t- \Delta t)}{2 \Delta t},
\end{equation*}
where ${\bm{x}}$ and ${\bm{u}}$ are the location and velocity of a particle, respectively;
$\Delta t$ is the time interval between the two consecutive frames.
The viscous term in the pressure gradient is computed using the least squares RBF-QR algorithm \citep{fornberg2011stable,larsson2013stable}.

Note that before computing the pressure gradients, we have the option to filter out the curl-free part in the corrupted velocity field using the divergence-free RBF-HHD solver.
We refer to this option as the velocity correction, which can improve the final reconstruction accuracy.
The tangential components of the velocities at the boundaries in the divergence-free part can be easily obtained by interpolating the velocimetry data.
\textcolor{black}{This divergence-free RBF-HHD solver and velocity correction is a convenient `add-on' to the curl-free pressure solver.
Other techniques could be used to achieve this goal as well.}

\begin{figure}[!htb]
	\centering
 \includegraphics[width=0.5\columnwidth]{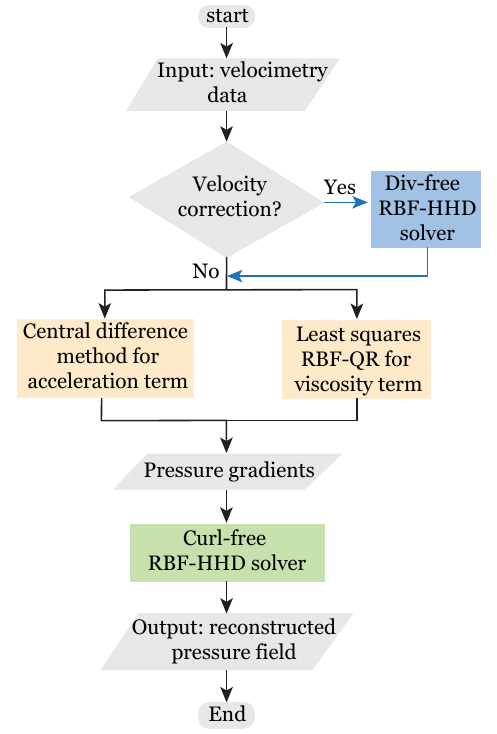}
	\caption{A flowchart illustrates the reconstruction process in the 2D cylinder flow validation test.}
 \label{fig: flowchart}
\end{figure}

After recovering the pressure gradients, we use the curl-free RBF-HHD solver to reconstruct the pressure fields.
We specify the tangential components of the pressure gradients at the boundaries in the curl-free part, which are interpolated using the gradients in the entire domain.
A flowchart about the reconstruction process is presented for illustration purposes (see Fig.~\ref{fig: flowchart}). 

\begin{figure}[!ht]
	\centering
\includegraphics[width=1\columnwidth]{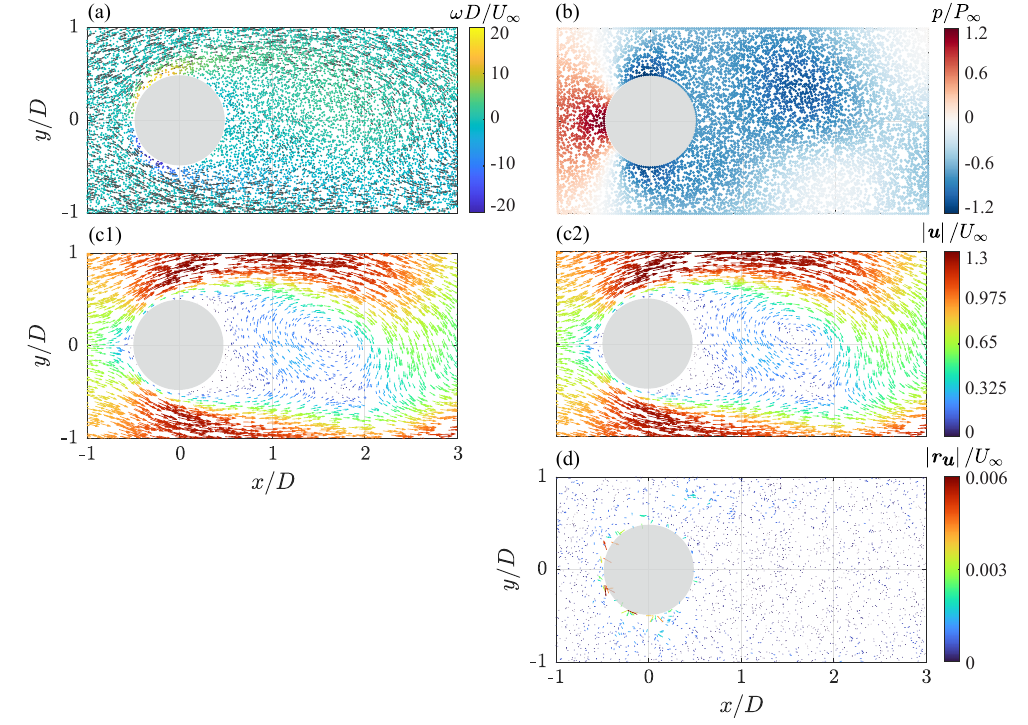}
	\caption{The true value and the synthetic data of the cylinder flow field. (a)~a quiver plot of the \textcolor{black}{true} velocity field overlaid on a \textcolor{black}{true} vorticity field; (b)~the true pressure field; (c1) and (c2):~corrupted velocity fields before and after the velocity correction, respectively; (d) the curl-free components removed from the velocity field (c1) by the divergence-free RBF-HHD solver.}
\label{fig: cylinder2d_output1}
\end{figure}

\begin{figure}[!ht]
	\centering
\includegraphics[width=1\columnwidth]{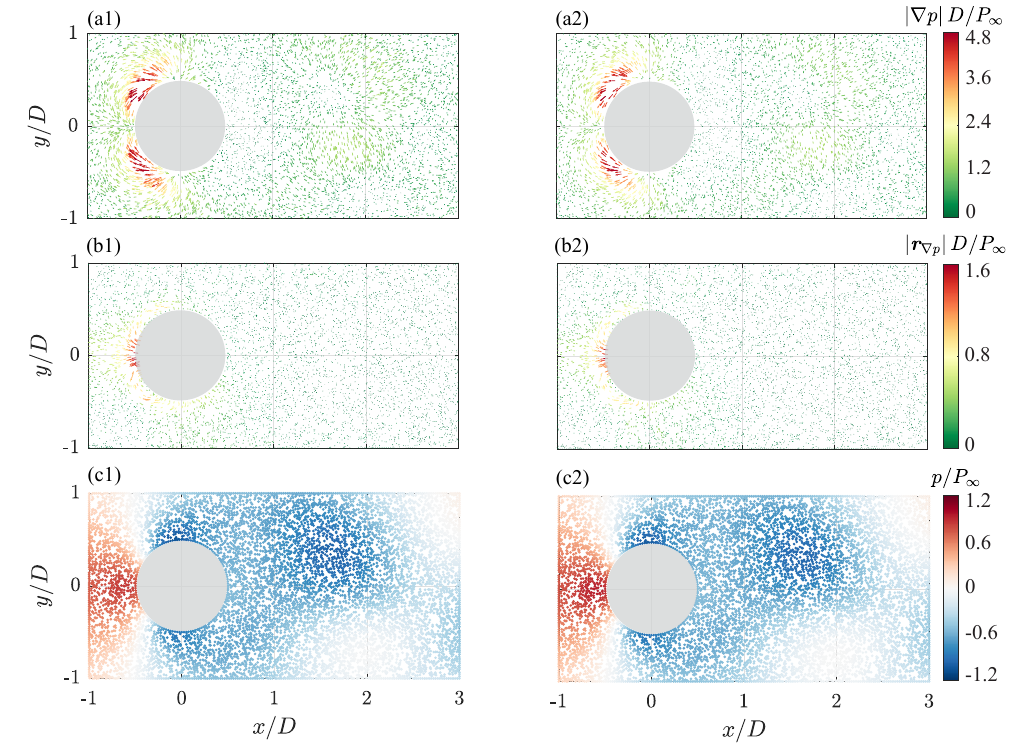}
	\caption{Reconstructed pressure gradients and pressure fields. (a1) and (a2): The reconstructed pressure gradients, computed from the data illustrated in Fig.~\ref{fig: cylinder2d_output1}(c1) and (c2), respectively. (b1) and (b2): The detected divergence-free bias in the pressure gradients of (a1) and (a2), respectively, identified by the curl-free RBF-HHD solver. (c1) and (c2): The reconstructed pressure fields after (b1) and (b2) are removed from (a1) and (a2), respectively.}
\label{fig: cylinder2d_output2}
\end{figure}

\begin{figure}[!htb]
	\centering
 \includegraphics[width=1\columnwidth]{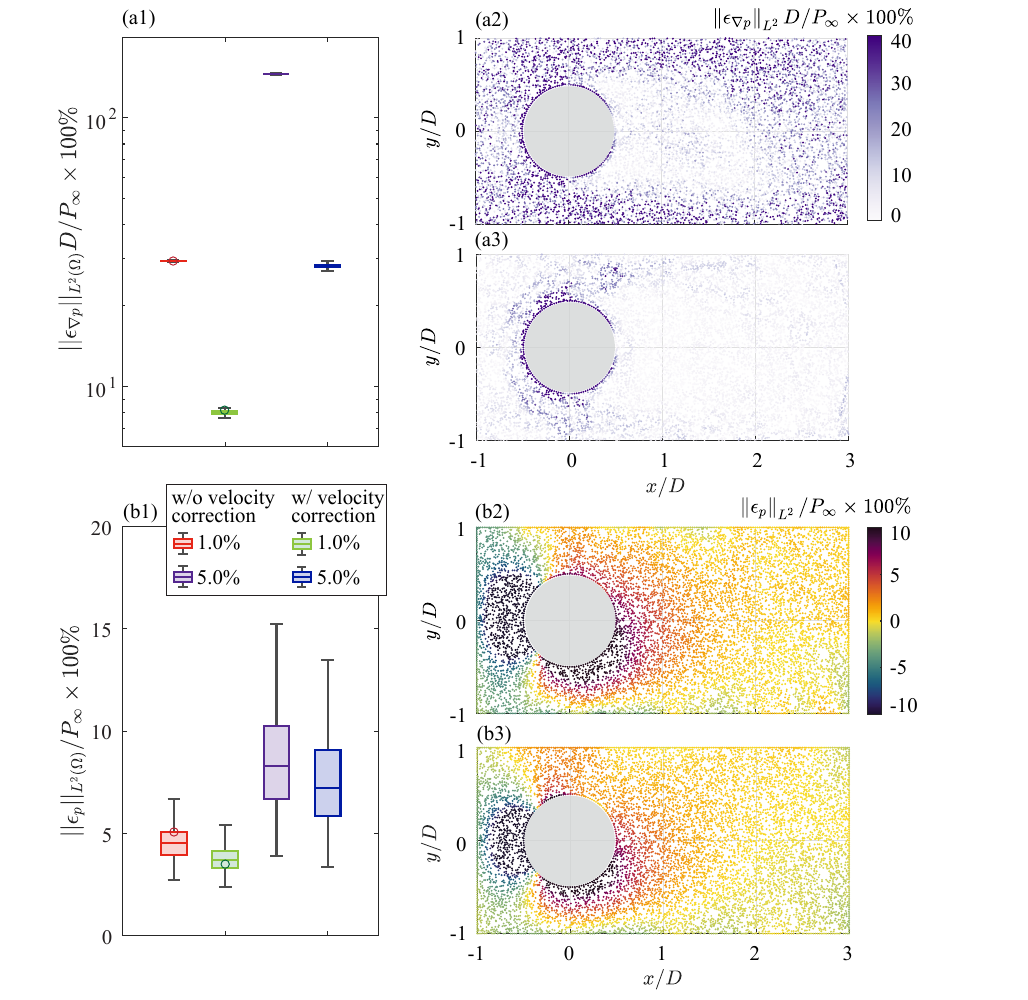}
	\caption{Statistics and typical results for the errors in the reconstructed pressure gradient and pressure fields. A box plot of the error in the reconstructed pressure gradient (a1) and pressure (b1) from 500 independent tests. Horizontal bars in the middle of the boxes show the median while the upper and lower edges of the box indicate the 25 and 75 percentiles. The upper and lower whiskers bound the 95\% confidence intervals of the error. The symbols within the boxes mark where the corresponding error is shown in (a2) \& (a3) for the reconstructed pressure gradients; and in (b2) \& (b3) for the reconstructed pressure field. (a2) \& (b2) and (a3) \& (b3) are error fields based on the reconstruction without and with the velocity correction, respectively.}
 \label{fig: error}
\end{figure}

Typical results of reconstruction and error statistics based on 500 independent tests are presented in Figs.~\ref{fig: cylinder2d_output1}~--~\ref{fig: error}.
Figure~\ref{fig: cylinder2d_output1} illustrates the true value of the velocity, vorticity, and pressure fields for reference (a and b), as well as the synthetic velocimetry data (c1) and the corrected velocity field (c2).
The synthetic velocimetry data are generated based on the 1\% noise.
The curl-free component (d) is removed from the contaminated velocity field (c1) and we obtain the divergence-free velocity field (c2).

Figure~\ref{fig: cylinder2d_output2} shows the reconstructed pressure gradients (a1)~--~(a2) and pressure fields (c1)~--~(c2) using the curl-free RBF-HHD solver.
The left and right columns in Fig.~\ref{fig: cylinder2d_output2} are the results based on the velocity field without and with the velocity correction, respectively.
The removed divergence-free bias is shown (b1)~--~(b2).
For this particular flow, the divergence-free error in the pressure field peaks near the stagnation region of the flow. 
This is perhaps due to intrinsically high-pressure gradients and relatively low particle density near the wall that conflicts with the boundary condition of the curl-free RBF-HHD solver.
Nevertheless, as shown in (c1)~--~(c2), the reconstructed pressure fields are similar to the ground truth (see Fig.~\ref{fig: cylinder2d_output1}(b)), even if the pressure gradients are contaminated. 

The errors in the reconstructed pressure gradients and pressure fields are shown in Fig.~\ref{fig: error}.
The synthetic data used in the tests were generated by adding 1\% and 5\% noise to the true value of the velocity field.
When the divergence-free correction was not applied to the velocity field, the median of the error in the reconstructed pressure field was about 4.6\% (red box in (b1)) for the 1\% noise level, despite the high error of about 29.5\% (red box in (a1)) which persisted in the pressure gradients.
Even if the velocity data were highly contaminated (5\% noise level), the median of the error in the reconstructed pressure fields only increased to 8.3\% (see purple box in (b1)).
This indicates that the curl-free RBF-HHD solver is robust to the noise in the data, even if the data are highly contaminated.

If the divergence-free velocity correction was employed, the errors in the reconstructed pressure gradients can be significantly reduced to about 8\% and 28\% (green and blue box in Fig.~\ref{fig: error}(a1)) for 1\% and 5\% noise levels in the velocity field, respectively. 
The corresponding median of the errors in the reconstructed pressure fields was reduced to about 3.7\% and 7.2\% (green and blue box in (b1)).
This error reduction was also evident when comparing the error field of the pressure gradient in (a2) with (a3) and comparing the pressure in (b2) with (b3). 
This demonstrates that the divergence-free RBF-HHD solver can further improve the pressure reconstruction.

Based on the above test, we want to highlight that the RBF-HHD solver is robust to errors in the input data.
It can filter out random noise and remove curl-free and divergence-free bias in the velocity and pressure gradient fields, respectively.
The RBF-HHD solver has no strict requirements on the data structure and the geometry of the domain, and it is suited for processing noisy and gappy PTV and/or PIV data.

\section{Conclusions}
\label{sect: conclusions}
In this study, we conduct an error propagation analysis concerning the general Pressure Gradient Integration (PGI) for reconstructing the pressure field from image velocimetry data \textcolor{black}{for incompressible flows}. 
Our analysis yields several findings that are of practical interest.

First, the upper bound of the error from PGI shows that applying the HHD-regularization on a corrupted pressure gradient field can significantly reduce the error in the reconstructed pressure.
This result is independent of the experimental techniques or the specific implementation of HHD.
The HHD can uniquely decompose a corrupted pressure gradient field into curl-free, which satisfies the Path Independence Property (PIP) of a pressure gradient field, and a divergence-free part, which should not be in the pressure gradient. 
If we remove the divergence-free part and reconstruct the pressure field solely based on the curl-free part of the corrupted pressure gradient, a pressure reconstruction with improved accuracy is expected.

Second, we argue that the HHD-regularized PGI can be considered a continuous limit of the ODI.
While the ODI endeavors to enforce the PIP of the contaminated pressure gradients, the HHD-regularized PGI can precisely remove divergence-free bias in the pressure gradients.

Third, our error estimates suggest that the HHD-regularized PGI can potentially outperform the pressure Poisson equation (PPE) solvers since the HHD-regularized PGI has a lower upper bound than that of the PPE, which is rooted in integration twice or once to obtain the pressure from $\nabla^2 p = f$ and $\nabla p = \bm{g}$, respectively.

Lastly, discrete Fourier analysis is applied to expose the error scaling laws for the PPE and PGI when the data are contaminated by point-wise zero-mean Gaussian noise. 
The errors in the pressure field are $\left \| \epsilon_p \right \|^{\text{PPE}}_{L^2(\Omega)} \sim L^2 \sigma_f N^{-1}$ and $\left \| \epsilon_p\right \|^{\text{PGI}}_{L^2(\Omega)} \sim L \sigma_{\bm{g}} N^{-1}$ for the PPE and PGI, respectively.
From the scaling laws, we find that 
i) high spatial resolution data (i.e., a sufficiently large $N$, which is the number of nodal points in one direction of the domain) can improve reconstruction quality; 
ii) both the PPE and PGI propagate the random error in the same way, and 
iii) the PPE and PGI amplify the error by the factors $L^2$ and $L$ respectively, where $L$ is the length scale of a square domain.

In this work, we propose to use RBF-HHD solvers, employing divergence/curl-free kernels for incompressible flows, to reconstruct pressure fields from image velocimetry data.
The RBF-HHD solvers can provide divergence-free and curl-free corrections to the velocity fields and pressure gradients, respectively.
This solver has several advantages \textcolor{black}{or features}:
i) flexible computation on scattered or structured data in a complex domain, requiring no Dirichlet boundary conditions except for reference pressure at a point,
ii) complete elimination of divergence/curl-free bias in measured data, ensuring accurate reconstruction;
iii) robustness against random noise due to its capability of adopting regression; and
iv) \textcolor{black}{enforcing divergence/curl-free constraints without using Lagrangian multipliers.}
Validation using synthetic PTV data of a 2D laminar cylinder flow and synthetic PIV data confirms the accuracy and robustness of our solvers.

\section{Acknowledgment}
We thank the Digital Research Alliances of Canada for providing high-performance computing resources. 
This work is partially supported by the Natural Sciences and Engineering Research Council of Canada (NSERC) Discovery Grant (RGPIN-2020-04486).  
G.B.W was partially supported by NSF grant DMS-1952674. J.P.W. was partially supported by NSF grant DMS-2206762.

\section{Appendices}
\appendix

\section{Discrete Fourier Analysis for PPE and PGI}
\label{sec: scaling laws for PPE and PGI apdx}

\subsection{Pressure Poisson Equation}
\label{sec: PPE scaling law apdx}
We consider a 2D domain of $ (x,y) \in \Omega = [0,L]\times [0,L]$ and define the 2D Fourier expansions for some quantity (e.g., $\epsilon(\bm{x}) = \epsilon(x,y)$) in this domain as 
\begin{equation}
\label{eq: Fourier nontruncated apdx}
    \epsilon(\bm{x}) = \sum_{k=-\infty}^{\infty}\sum_{l=-\infty}^{\infty} \hat{\epsilon} [k,l]e^{j\frac{2\pi}{L}(kx + ly)}  = \sum_{\bm{k}} \hat{\epsilon} [\bm{k}]e^{j\frac{2\pi}{L}\bm{k} \cdot \bm{x}},
\end{equation}
where $\hat{\epsilon} [\bm{k}] = \hat{\epsilon} [k,l]$ is the Fourier coefficient.
The domain $\Omega$ can be discretized into an $N \times N$ grid with a uniform grid spacing $h=L/(N-1)$. 
Sampling $\epsilon(\bm{x})$ at the \textcolor{black}{nodal points $\bm{x} = (mh,nh) = \bm{m}h,~\text{where}~m,n = 0,1,2,...,N$} leads to truncating the infinite summation in \eqref{eq: Fourier nontruncated apdx} into a finite discretized one \citep{iserles2009first}:
\textcolor{black}{
\begin{equation}
\begin{aligned}
\label{eq: Fourier truncated apdx}
    \epsilon(\bm{x}) = \sum_{k=0}^{N-1}\sum_{l=0}^{N-1}\hat{\epsilon} [k,l]e^{j \frac{2\pi}{L} (kx + ly)}  &= \sum_{\bm{k}}\hat{\epsilon} [\bm{k}]e^{j \frac{2\pi}{L} \bm{k}\cdot \bm{x}}  \\
    &= \sum_{\bm{k}} \hat{\epsilon} [\bm{k}]e^{j \frac{2\pi }{(N-1)} \bm{k}\cdot \bm{m}}  = \epsilon(\bm{m}). 
\end{aligned}
\end{equation}
}

Let $\epsilon$ in \eqref{eq: Fourier truncated apdx} be $\epsilon_f$ or $\epsilon_p$, and invoking $\nabla^2 \epsilon_p = \epsilon_f$ yields the relationship between the Fourier coefficients for $\epsilon_f$ and $\epsilon_p$: 
\begin{align}
\label{eq: hat lamda hat apdx}
\hat\epsilon_p[\bm{k}] = - \frac{L^2}{4\pi^2 |\bm{k}|^2} \hat\epsilon_f[k,l] = \lambda_{\bm{k}}^{-1}\hat \epsilon_f[\bm{k}],
\end{align}
where the coefficients are specified by $\lambda_{\bm{k}}^{-1} = - \frac{L^2}{4\pi^2 |\bm{k}|^2} = - c^2 \frac{L^2}{|\bm{k}|^2}$.  The constant $c^{-1} = 2\pi$ may vary depending on the specific geometry of the domain and the boundary condition setup (different boundary conditions may require a different basis than the Fourier series here, but the same general behavior will still dominate). 

We suppose that $\epsilon_f(\bm{m})$ represents point-wise independent zero mean Gaussian noise with a constant variance $\sigma_{\epsilon_f}^2$, i.e., the noise is independently added at all of the grid points in the domain.  Then its Fourier coefficients are also zero mean Gaussian noise \citep{rice1944mathematical}:
\begin{equation}
\label{eq: distribution epsilon_f hat apdx}
    {\hat{\epsilon}}_f(\bm{m}) \sim \mathcal{N} \left ( \begin{bmatrix}
0 \\ 
0
\end{bmatrix},
\begin{bmatrix}
\sigma_{\epsilon_f}^2 / N^2 & 0 \\ 
0 & \sigma_{\epsilon_f}^2 / N^2
\end{bmatrix} \right ),
\end{equation}
and the expected value of its square is  $\mathbb{E} \left [ \hat{\epsilon}_f^2 \right ] = {\sigma_{\epsilon_f}^2}/{N^2}$.

\textcolor{black}{By the Fourier-Plancherel Theorem,
\begin{align}
    \sum_{\bm{m}} \left |\epsilon_p(\bm{m}) \right |^2 = \sum_{\bm{k}} \left |\hat{\epsilon}_p[\bm{k}] \right |^2 = \sum_{\bm{k}}\lambda_{\bm{k}}^{-2} \left |\hat{\epsilon}_f[\bm{k}] \right |^2.
\end{align}
Taking the expectation of both sides, we find that
\begin{align}
    \mathbb{E}\left[ \sum_{\bm{m}}|\epsilon_p(\bm{m})|^2\right] &= \mathbb{E}\left[\sum_{\bm{k}}\lambda_{\bm{k}}^{-2}|\hat{\epsilon}_f[\bm{k}]|^2\right]\\
    &=\sum_{\bm{k}}\lambda_{\bm{k}}^{-2}\mathbb{E}\left[|\hat{\epsilon}_f[\bm{k}]|^2\right]\\
    &=\frac{L^4 \sigma_{\epsilon_f}^2}{16\pi^4 N^2}\sum_{\bm{k}}|\bm{k}|^{-4} = c^4\frac{L^4\sigma_{\epsilon_f}^2}{N^2} \mathcal{P}^2(N),
\end{align}
where $\mathcal{P}^2(N) = \sum_{\bm{k}}|\bm{k}|^{-4}$ can be directly computed given a certain $N$.}
\textcolor{black}{Note that the expected value of the summation above is an approximation of the $L^2$ norm across the entire domain so that
\begin{equation}
    \|\epsilon_p\|_{L^2(\Omega)} \approx \sqrt{\mathbb{E}\left[\sum_{\bm{m}}|\epsilon_p(\bm{m})|^2\right]} = c^2\frac{L^2\sigma_{\epsilon_f}}{N}\mathcal{P}(N).
\end{equation}}

\begin{figure}[!h]
	\centering
	\includegraphics[width=0.5\columnwidth]{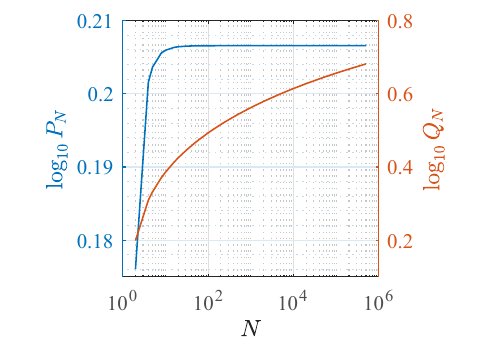}
	\caption{Double sums $\mathcal{P}(N)$ \& $\mathcal{Q}(N)$ vs. the number of nodal points in one direction ($N$).}
\label{fig: doubleSum}
\end{figure}

Note the value of $c^2 = \pi^{-2}/4$ is rooted in the periodic boundary condition that is assumed by the above Fourier analysis. 
Other boundary conditions and changes in the geometry of the domain will change the specific value of $c^2$ via an alternative choice of an orthogonal basis (as opposed to the Fourier basis used here). 
However, the power law respecting $\sigma_{\epsilon_f}$, $L$, and $N$ will not be altered.
Interestingly, for a sufficiently large $N$, $\mathcal{P}(N)$ grows very slowly with $N$, and $ \mathcal{P}(N) \sim N^0$ is almost a constant (see Fig.~\ref{fig: doubleSum}). 
Thus, a simple approximation of the scaling law for the error in the pressure field is given as 
\begin{equation}
    \begin{aligned}
    \label{eq: expected error PPE apdx}
    || \epsilon_p||_{L^2(\Omega)} \sim 
  L^2 \sigma_{\epsilon_f} N^{-1},
    \end{aligned}
\end{equation}
for a large $N$, and this is the main result of \eqref{eq: scaling law PPE simple main text} listed in Sect.~\ref{Sec: Error scaling for PPE and PGI}. 

We can also estimate the variance of $\epsilon_p$ as 
\begin{equation}
\label{eq: var error PPE apdx}
 \sigma_{\epsilon_p}^2 =    \mathbb{V} \left[ \epsilon_p \right] = \mathbb{E}[\epsilon_p^2] - \mathbb{E}[\epsilon_p] ^2 \leq \mathbb{E}[\epsilon_p^2] = c^4 \frac{L^4 \sigma_{\epsilon_f}^2}{ N^2}  \mathcal{P}^2(N) \sim {L^4 \sigma_{\epsilon_f}^2}{N^{-2}}.
\end{equation}
Normalizing $\sigma_{\epsilon_p}$ by $\sigma_{\epsilon_f}$ and rearranging \eqref{eq: var error PPE apdx} leads to 
\begin{equation}
\label{eq: normalzied var PPE apdx}
     \sigma_{\epsilon_p}/\sigma_{\epsilon_f} \sim L^2 N^{-1}.
\end{equation}
Equations \eqref{eq: expected error PPE apdx} and \eqref{eq: normalzied var PPE apdx} together indicate that data at high spatial resolution (e.g., raising $N$) will be able to attenuate random noise in the data $f$, in terms of reducing the expectation and standard deviation of the error, with the same roll-off of $N^{-1}$.

\subsection{Direct Pressure Gradient Integral}
\label{sec: PGI scaling law apdx}

\textcolor{black}{
Similar to the \textcolor{black}{analysis for} PPE, we develop a scaling law for error propagation in the PGI. 
The Fourier coefficients for the error in the pressure is  
\begin{equation*}
   j \frac{2\pi}{L} \bm{k}\cdot\hat{\epsilon}_p[\bm{k}] = \hat{\epsilon}_{\nabla p}[\bm{k}],
\end{equation*}
indicating that
\begin{align}
\label{eq: hat lamda hat odi apdx}
|\hat\epsilon_p[\bm{k}]| =  \frac{L}{2\pi |\bm{k}|} | \hat\epsilon_{\nabla p}[k,l] |= \lambda_{\bm{k}}^{-1} | \hat \epsilon_{\nabla p}[\bm{k}] |,
\end{align}
where the coefficients $\lambda_{\bm{k}} =  \frac{L}{2\pi |\bm{k}|} =  c \frac{L}{|\bm{k}|}$, and $c^{-1} = 2\pi$ is a constant that may vary depending on the geometry of the domain and the specific boundary conditions.}

Since $\epsilon_{\nabla p}
(\bm{m})$ is assumed to be a point-wise independent zero mean Gaussian noise with a constant variance $\sigma_{\epsilon_{\nabla p}}^2$ in each direction, then $ | \hat{\epsilon}_{ \nabla p } | \sim \mathcal{R}(\frac{\sigma}{N})$, where $\mathcal{R}$ is a Rayleigh distribution \citep{chattamvelli2022continuous}. 
Utilizing the Fourier-Plancherel Theorem leads to
\begin{align}
\label{eq: squared epsilon p odi apdx}
    \sum_{\bm{m}}|\epsilon_p(\bm{m})|^2 = \sum_{\bm{k}}|\hat{\epsilon}_p [\bm{k}]|^2 = \sum_{\bm{k}} \lambda_{\bm{k}}^{-2} | \hat{\epsilon}_{\nabla p} [\bm{k}] |^2.
\end{align}

Computing the expectation for both sides of \eqref{eq: squared epsilon p odi apdx} and invoking the 
the expectation of a Rayleigh distributed variable (i.e., $\mathbb{E}[| \hat{\epsilon}_{ \nabla p } |] = \sqrt{\frac{\pi}{2}} \frac{\sigma}{N}$ and  $\mathbb{E} \left [ \| \hat{\epsilon}_{\nabla p} \| ^2 \right ] = \frac{\pi \sigma^2}{2 N^2}$), we arrive at 
\begin{equation}
\label{eq: error squared dpgi}
    \begin{aligned}
    \mathbb{E} \left [\sum_{\bm{m}}|\epsilon_p(\bm{m})|^2 \right ] & = \mathbb{E} \left [ \sum_{\bm{k}} \lambda_{\bm{k}}^{-2} | \hat{\epsilon}_{\nabla p} [\bm{k}] |^2 \right ] \\
    & = \sum_{\bm{k}} \mathbb{E} \left[  \lambda_{\bm{k}}^{-2} | \hat{\epsilon}_{\nabla p} [\bm{k}] |^2 \right]\\
    & =   \frac{L^2 \sigma^2}{8 \pi N^2}  \sum_{\bm{k}} {|\bm{k}|^{-2}} = c^2 \frac{L^2 \sigma^2}{ N^2} \mathcal{Q}^2(N), 
    \end{aligned}
\end{equation}
where $\mathcal{Q}^2(N) = \sum_{\bm{k}} {|\bm{k}|^{-2}}$ can be directly evaluated given a certain $N$. 

Taking the square root on both sides of \eqref{eq: error squared dpgi}, we can estimate the error in the reconstructed pressure
\begin{equation}
\label{eq: scaling law dpgi}
    \begin{aligned}
    \| \epsilon_p\|_{L^2(\Omega)} \approx 
\sqrt{\mathbb{E} \left[\sum_{\bm{m}} |\epsilon_p(\bm{m})|^2 \right]} 
 = c \frac{L \sigma}{N} \mathcal{Q}(N), 
    \end{aligned}
\end{equation}
where $c^2 = \pi^{-1}/8$ is based on the periodic boundary condition that is implied by the Fourier analysis. 
Similar to the PPE analysis, for a sufficiently large $N$, $\mathcal{Q}(N)$ grows, but slowly with $N$, and $ \mathcal{Q}(N) \sim N^0$ is almost a constant as well (see Fig.~\ref{fig: doubleSum}). 
Thus, an approximation of the scaling law for the error propagation for the PGI is
\begin{equation}
\label{eq: scaling law dpgi simple apdx}
    \begin{aligned}
    || \epsilon_p||_{L^2(\Omega)} \sim 
  L \sigma N^{-1},
    \end{aligned}
\end{equation}
which is the result of \eqref{eq: scaling law dpgi simple main text} in the main text.
We can also estimate the variance of $\epsilon_p$ as 
$$\sigma_{\epsilon_p}^2  = \mathbb{V} \left[ \epsilon_p \right] = \mathbb{E}[\epsilon_p^2] - \mathbb{E}[\epsilon_p] ^2 \leq \mathbb{E}[\epsilon_p^2] = c^2 \frac{L^2 \sigma^2}{ N^2}  \mathcal{Q}(N) \sim c^2 {L^2 \sigma^2}{N^{-2}}.$$
Rearranging the above equation leads to 
\begin{equation}
\label{eq: sigma PGI scaling apdx}
    \frac{\sigma_{\epsilon_p}}{L\sigma} \sim N^{-1}. 
\end{equation}
Note, if the slight growth of $\mathcal{Q}(N)$ is considered, the power of $N$ in \eqref{eq: sigma PGI scaling apdx} is slightly larger than $-1$, and is consistent with the experiments in \citet{wang2023green}.

\section{Pressure Solvers}
This section reports the implementation details of the pressure solvers used in this work.

\label{sect: solvers}
\subsection{Finite difference based solvers}
\label{sect: solvers fdm}

We discretize the pressure field on a uniform 2D Cartesian mesh, with a uniform grid spacing $h$ in $x$ and $y$ directions. 
Central difference based Poisson solvers with different order accuracy are implemented. 

The second-order central finite difference scheme for a 2D discretized PPE is given by:
\begin{equation}
    4p_{i,j}-p_{i+1,j}-p_{i-1,j} - p_{i,j+1}-p_{i-1,j} = -h^2 f_{i,j},
    \label{eq: 2nd ppe}
\end{equation}
where $p_{i,j}$ is the pressure at the interior node $(i,j)$, $f_{i,j}$ is the source term at the interior node $(i,j)$, $i,j = 2,3, \dots, N-1$.
Similarly, the fourth-order central finite difference stencil is given by:
\begin{equation}
\begin{split}
    60p_{i,j}+p_{i+2,j}-16p_{i+1,j}-16p_{i-1,j}+p_{i-2,j}+p_{i,j+2}-16p_{i,j+1}-16p_{i-1,j}+p_{i-2,j} \\
    = -{12h^2} f_{i,j}.
    \label{eq: 4th ppe}
\end{split}
\end{equation}
Neumann conditions are implemented using \textcolor{black}{a first-order finite difference scheme} on the boundary. 
The resulting linear system is solved using MATLAB's \verb+mldivide+ solver.




\subsection{RBF-HHD solver}
\label{sect: solvers hhd}
The RBF-HHD solver is based on the work by \citet{fuselier2016high,fuselier2017radial}. 
\textcolor{black}{We follow similar notations to those in their works to facilitate consistent reference for readers.}

\subsubsection{\textcolor{black}{RBF kernels}}
A scalar field $f$ sampled at the points $X=\{x_1,\ldots,x_N\}\subset\mathbb{R}^d$ can be approximated using a linear combination of positive \textcolor{black}{definite radial kernels $\phi: \mathbb{R}^d \times \mathbb{R}^d \to \mathbb{R}$ as}
\begin{equation}
    s_f = \sum^{N}_{j=1}\phi(\cdot,x_j)c_j,
    \label{eq: hhd scalar value}
\end{equation}
where $\phi(\cdot,x_j) = \phi(|\cdot-x_j|)$ and the expansion coefficients $c_j$ are determined by forcing the interpolant to coincide with the given data $ \left . f \right|_{X}$, i.e., $\left . s_f \right |_{X} = \left . f \right|_{X}$.
\textcolor{black}{This idea can be extended to samples of a vector field $\mathbf{f}: \mathbb{R}^d \to \mathbb{R}^d$ using positive definite matrix-valued kernels ${\bf{\Phi}}: \mathbb{R}^d \times \mathbb{R}^d \to \mathbb{R}^d \times \mathbb{R}^d$ (see the next subsection for examples).  The vector-field approximant takes the form}
\begin{equation}
    {\bf{s}_f} = \sum^{N}_{j=1}{\bf{\Phi}}(\cdot,x_j){\bf{c}}_j,\label{eq:hhd_vec_value}
\end{equation}
\textcolor{black}{where the 
expansion coefficients are $d$ dimensional vectors, i.e., ${\bf{c}}_j\in\mathbb{R}^d$.  
These coefficients can be determined by enforcing interpolation $\mathbf{s}_{\mathbf{f}}\bigr |_X = \mathbf{f} \bigr |_X$, which leads to the linear system}
\begin{equation}
\underbrace{
    \begin{bmatrix}
    {\bf{\Phi}}(x_1,x_1) & \hdots & {\bf{\Phi}}(x_1,x_N) \\
    \vdots & \ddots & \vdots \\
    {\bf{\Phi}}(x_N,x_1) & \hdots & {\bf{\Phi}}(x_N,x_N) \\
    \end{bmatrix}}_{\displaystyle A}
\underbrace{
    \begin{bmatrix}
     {\bf{c}}_1 \\ 
     \vdots \\ 
     {\bf{c}}_N
    \end{bmatrix}}_{\displaystyle \underline{\bf{c}}} = 
\underbrace{
    \begin{bmatrix}
    {\bf{f}}_1 \\ 
    \vdots \\ 
    {\bf{f}}_N
    \end{bmatrix}}_{\displaystyle \underline{\bf{f}}},
    \label{eq: hhd linear system}
\end{equation}
\textcolor{black}{where $A$ is an $Nd \times Nd$ matrix-valued {Gram} matrix, $\underline{\bf{c}}$ is a column vector of size $Nd$ that needs to be computed, and $\underline{\bf{f}}$ is a column vector of size $Nd$ containing the samples of the vector field $\bf{f}$.}

\textcolor{black}{To further elaborate on the structure of the system \eqref{eq: hhd linear system}, we consider the case of $d=2$. Denoting the entries of ${\bf\Phi}$ as ${\bf\Phi}_{11}$, ${\bf\Phi}_{12}$, ${\bf\Phi}_{21}$, and ${\bf\Phi}_{22}$, the matrix $A$ has the form
\begin{equation}
A = 
    \begin{bmatrix}
    {\bf\Phi}_{11}(x_1,x_1) & {\bf\Phi}_{12}(x_1,x_1) & \hdots & {\bf\Phi}_{11}(x_1,x_N) & {\bf\Phi}_{12}(x_1,x_N) \\
    {\bf\Phi}_{21}(x_1,x_1) & {\bf\Phi}_{22}(x_1,x_1) &  \hdots & {\bf\Phi}_{21}(x_1,x_N) & {\bf\Phi}_{22}(x_1,x_N) \\
    \vdots & \vdots  & \ddots & \vdots & \vdots \\
    {\bf\Phi}_{11}(x_N,x_1) & {\bf\Phi}_{12}(x_N,x_1) &  \hdots & {\bf\Phi}_{11}(x_N,x_N) & {\bf\Phi}_{12}(x_N,x_N) \\
    {\bf\Phi}_{21}(x_N,x_1) & {\bf\Phi}_{22}(x_N,x_1) & \hdots & {\bf\Phi}_{21}(x_N,x_N) & {\bf\Phi}_{22}(x_N,x_N)
    \end{bmatrix}.
    \label{eq: entries of A}
\end{equation}
Also, by denoting the entries of ${\bf{c}}_j$ and ${\bf{f}}_j$ as ${\bf{c}}_j = \begin{bmatrix} c_{1,j} & c_{2,j}\end{bmatrix}^{\intercal}$ and  ${\bf{f}}_j = \begin{bmatrix} f_{1}(x_j) & f_{2}(x_j)\end{bmatrix}^{\intercal}$, the vectors $\underline{\bf{c}}$ and $\underline{\bf{f}}$ are given as 
\begin{align*}
\underline{\bf{c}} &= \begin{bmatrix} c_{1,1} & c_{2,1} & c_{1,2} & c_{2,2} & \hdots & c_{1,N} & c_{2,N} \end{bmatrix}^{\intercal}, \\
\underline{\bf{f}} &= \begin{bmatrix} f_{1}(x_1) & f_2(x_1) & f_1(x_2) & f_{2}(x_2) & \hdots & f_1(x_N) & f_2(x_N) \end{bmatrix}^{\intercal}. 
\end{align*}
}

\subsubsection{\textcolor{black}{Divergence/curl-free RBF-HHD kernels}}
\textcolor{black}{Divergence-free and curl-free interpolants in $d=2$ and 3 dimensions can be constructed within the generalized matrix-valued interpolation framework outlined above.}

\textcolor{black}{
For a divergence-free interpolant, the kernel ${\bf \Phi}^{\text{df}}$ given in \eqref{eq:div_curl_kernels} needs to be used in \eqref{eq:hhd_vec_value}, while for a curl-free interpolant one would similarly use the kernel ${\bf \Phi}^{\text{cf}}$ in \eqref{eq:div_curl_kernels}.}
\textcolor{black}{For a bounded domain $\Omega\in\mathbb{R}^d$, and $x,y\in\Omega$, both of these kernels can be written explicitly by applying the derivative operators in \eqref{eq:div_curl_kernels} to $\phi(|x-y|)$ and simplifying.  The resulting formulas are given as
\begin{align}
{\bf{\Phi}}^\text{df}(x,y) &= -(r^2\zeta(r) + d \chi(r))\mathbf{I} + \zeta(r)(x-y)(x-y)^{\intercal}, \label{eq: entries of div kernel} \\
{\bf{\Phi}}^\text{cf}(x,y) &=  -\zeta(r)(x-y)(x-y)^{\intercal},  \label{eq: entries of curl-free kernel}
\end{align}
where $\mathbf{I}$ is the $d$-by-$d$ identity matrix, $r=|x-y|$, $\chi(r) = \frac{1}{r}\phi'(r)$, and $\zeta(r) = \frac{1}{r}\chi'(r)$. Here $y$ is thought of as the `center' of the matrix-valued kernel.}

\textcolor{black}{
While these kernels can be used separately in \eqref{eq:hhd_vec_value}, we will combine them as in \eqref{eq:div_curl_kernels} to mimic the HHD.  Using \eqref{eq: entries of div kernel} and \eqref{eq: entries of curl-free kernel}, the combined kernel simplifies to
\begin{equation} 
\mathbf{\Phi}(x,y) = -(r^2\zeta(r) + d\, \chi(r))\mathbf{I}. \label{eq:Phi_decomp}
\end{equation}
This is a diagonal kernel, meaning it only has non-zero entries on its diagonal.  In this case, the Gram matrix \eqref{eq: entries of A} simplifies so that all entries involving ${\bf\Phi}_{12}$ and ${\bf\Phi}_{21}$ are zero.} 

\textcolor{black}{We note that when constructing ${\bf \Phi}^{\text{df}}$ and ${\bf \Phi}^{\text{cf}}$ from the scalar radial kernel $\phi$ such as the popular Gaussian, multiquadric, or Mat\'ern~\citep{fasshauer2007meshfree}, they are symmetric and positive definite~\citep{fuselier2016high}.  This means that the matrix $A$ in \eqref{eq: entries of A} is symmetric positive definite so that the linear system \eqref{eq: hhd linear system} is uniquely solvable.
}

\subsubsection{\textcolor{black}{Decomposition based on divergence-free boundary conditions}}
Following~\citet{fuselier2017radial}, \textcolor{black}{boundary conditions for the normal component of the divergence-free part of the field} can be incorporated into the vector interpolant by modifying $\bf{s}_f$ as
\begin{equation}
        {\bf{s}_f} = \sum^N_{j=1} {\bf{\Phi}}(\cdot,x_j){\bf{c}}_j + \sum^M_{i=1} {\bf{\Phi}}^{\text{df}}(\cdot,y_i){\bf{n}}_{y_i}{d}_i,
\label{eq: core equation div-free}
\end{equation}
where $X=\{x_1,\ldots,x_N\}$ are in the interior of the domain $\Omega$ and $Y=\{y_1,\ldots,y_M\}$ are on the boundary $\partial \Omega$;
${\bf{n}}_{y_i}$ is the unit outward normal vector on $\partial\Omega$ at $y_i$ and ${d}_i$ are another set of unknown (scalar) expansion coefficients.
Both sets of coefficients ${\bf{c}}_j$ and ${d}_i$ can be obtained by enforcing interpolation, which leads to the linear system of equations
\begin{equation}
    \renewcommand{\arraystretch}{1.1}
    \begin{bmatrix}
    A & B \\
    B^\intercal & C
    \end{bmatrix}
    \begin{bmatrix}
    \underline{\bf{c}} \\
    \underline{d}
    \end{bmatrix} = 
    \begin{bmatrix}
    \underline{\bf{f}} \\
    \underline{g}
    \end{bmatrix},
    \label{eq: div-free linear system}
\end{equation}
\textcolor{black}{where $A$, $\underline{\bf{c}}$, and $\underline{\bf{f}}$ are given by \eqref{eq: hhd linear system}, $\underline{g}$ is the vector containing samples of the scalar boundary condition on the normal component of the divergence-free part of the field at $y_i$, and $\underline{d}$ is the vector containing the scalar coefficients $d_i$}. \textcolor{black}{The $Nd \times M$ matrix $B$ is given as}
\renewcommand{\arraystretch}{1}
\begin{equation*}
    B = \begin{bmatrix}
        {\bf{\Phi}}^{\text{df}}(x_1,y_1) {\bf{n}}_{y_1} & \hdots & {\bf{\Phi}}^{\text{df}}(x_1,y_M) {\bf{n}}_{y_M} \\
        \vdots & \ddots & \vdots \\
        {\bf{\Phi}}^{\text{df}}(x_N,y_1) {\bf{n}}_{y_1} & \hdots & {\bf{\Phi}}^{\text{df}}(x_N,y_M) {\bf{n}}_{y_M}
    \end{bmatrix},
\end{equation*}
 \textcolor{black}{while the $M$-by-$M$ symmetric matrix $C$ has the entries $C_{ij} = {\bf{n}}^\intercal_{y_i} {\bf{\Phi}}^{\text{df}}(y_i,y_j) {\bf{n}}_{y_j}$.}
\textcolor{black}{Similar to \eqref{eq: entries of A}, when $d=2$, the entry ${\bf{\Phi}}^{\text{df}}(x_i,y_j) {\bf{n}}_{y_j}$ is a column vector of length two whose entries can be computed using matrix-vector multiplication with ${\bf{\Phi}}^{\text{df}}$ defined in \eqref{eq: entries of div kernel} with the 2D normal vector ${\bf{n}}_{y_j}$.}

\textcolor{black}{Once ${\bf{c}}_j$ and ${d}_i$ are solved for using \eqref{eq: div-free linear system}, we substitute them into \eqref{eq: core equation div-free} and use the decomposition ${\bf \Phi} = {\bf \Phi}^{\text{df}} + {\bf \Phi}^{\text{cf}}$ to obtain an approximate HHD of $\mathbf{f}$:} 
\begin{equation}
        {\bf{s}_f} = \underbrace{\sum^N_{j=1} {\bf{\Phi}}^\text{df}(\cdot,x_j){\bf{c}}_j + \sum^M_{i=1} {\bf{\Phi}}^\text{df}(\cdot,y_i){\bf{n}}_{y_i}{d}_i}_{\displaystyle\mathbf{s}^{\rm df}_\mathbf{f}} + \underbrace{\sum^N_{j=1} {\bf{\Phi}}^{\text{cf}}(\cdot,x_j){\bf{c}}_j}_{\displaystyle\mathbf{s}^{\rm cf}_\mathbf{f}},
        \label{eq:velocity_correction}
\end{equation}
where $\mathbf{s}^{\rm df}_\mathbf{f}$ and $\mathbf{s}^{\rm cf}_\mathbf{f}$ are the divergence-free and curl-free parts of ${\bf{s}_f}$, respectively.
A potential $\bm{\Psi}_{\bf{s}_{f}}$ for ${\bf s}_{\bf f}^{\text{df}}$ and a (scalar) potential $q_{\bf{s}_{f}}$ for ${\bf s}_{\bf f}^{\text{cf}}$ can also be computed \textcolor{black}{from ${\bf{c}}_j$ and ${d}_i$ using the formulas for ${\bf\Phi}^{\text{df}}$ and ${\bf\Phi}^{\text{cf}}$ given in \eqref{eq:div_curl_kernels3d}:}
\begin{align*}
{\bf{s}_f}^{\text{df}} &= \textbf{curl}\underbrace{\left(-\sum^N_{j=1} \textbf{curl} (\phi(|\cdot - x_j|) {\bf{c}}_j) - \sum^M_{i=1} \textbf{curl} (\phi(|\cdot - y_i|) {\bf{n}}_{y_i} ){d}_i\right)}_{\displaystyle := \bm{\Psi}_{\bf{s}_{f}}}, \\
{\bf{s}_f}^{\text{cf}} &= \nabla\underbrace{\left(-\sum^N_{j = 1} [\nabla \phi(|\cdot-x_j|)]^{\intercal} {\bf{c}}_j\right)}_{\displaystyle := q_{\bf{s}_{f}}},
\end{align*}
where $\bm{\Psi}_{\bf{s}_{f}}$ will be a scalar potential in 2D and a vector potential in 3D.
In this work, we use \eqref{eq:velocity_correction} for the velocity correction solver.

\subsubsection{\textcolor{black}{Decomposition based on curl-free boundary conditions}}
Alternatively, boundary conditions on the tangential component of the curl-free part of the field can be incorporated into $\bf{s}_f$ as
\begin{equation}
    {\bf{s}_f} = \sum^N_{j=1} {\bf{\Phi}}(\cdot,x_j){\bf{c}}_j + \sum^M_{i=1} {\bf{\Phi}}^{\text{cf}}(\cdot,y_i){\bf{t}}_{y_i}{d}_i,
    \label{eq: equation curl-free}
\end{equation}
where ${\bf{t}}_{y_i}$ are the unit tangential vectors to the boundary $\partial\Omega$ at $y_i$.  Note that in 2D, there is only one tangential direction, but in 3D there are two that must be included for each $y_i$.  For simplicity, we will focus on the 2D case.  Imposing interpolation conditions to determine the coefficients ${\bf{c}}_j$ and ${d}_i$ leads to the linear system
\begin{equation}
    \begin{bmatrix}
    A & B \\
    B^{\text{T}} & C
    \end{bmatrix}
    \begin{bmatrix}
    {\underline{\bf{c}}} \\
    {\underline{d}}
    \end{bmatrix} = 
    \begin{bmatrix}
    \underline{\bf{f}} \\
    \underline{g}
    \end{bmatrix},
    \label{eq: curl-free linear system}
\end{equation}
where $A$ is given by the \eqref{eq: hhd linear system}, $B$ can be written as
\begin{equation*}
    B = \begin{bmatrix}
        {\bf{\Phi}}^{\text{cf}}(x_1,y_1) {\bf{t}}_{y_1} & \hdots & {\bf{\Phi}}^{\text{cf}}(x_1,y_M) {\bf{t}}_{y_M} \\
        \vdots & \ddots & \vdots \\
        {\bf{\Phi}}^{\text{cf}}(x_N,y_1) {\bf{t}}_{y_1} & \hdots & {\bf{\Phi}}^{\text{cf}}(x_N,y_M) {\bf{t}}_{y_M}
    \end{bmatrix}.
\end{equation*}
$C$ is symmetric with entries $C_{ij} = {\bf{t}}^\intercal_{y_i} {\bf{\Phi}}^{\text{cf}}(y_i,y_j) {\bf{t}}_{y_j}$, and $\underline{g}$ is the vector containing samples of the scalar boundary condition on the tangential component of the curl-free part of the field at $y_i$.

\textcolor{black}{Once ${\bf{c}}_j$ and ${d}_i$ are solved for using} \eqref{eq: curl-free linear system}, we can substitute them into \eqref{eq: equation curl-free} to obtain the following approximate HHD of $\mathbf{f}$ 
\begin{equation*}
        {\bf{s}_f} = \underbrace{\sum^N_{j=1} {\bf{\Phi}}^\text{df}(\cdot,x_j){\bf{c}}_j}_{\displaystyle\mathbf{s}^{\rm df}_\mathbf{f}} + \underbrace{\sum^N_{j=1} {\bf{\Phi}}^\text{cf}(\cdot,x_j){\bf{c}}_j + \sum^M_{i=1} {\bf{\Phi}}^{\text{cf}}(\cdot,y_j){\bf{t}}_{yj}{d}_j}_{\displaystyle\mathbf{s}^{\rm cf}_\mathbf{f}}.
\end{equation*}
Recovering the potentials for ${\bf{s}_f}$ for the curl-free boundary conditions is similar to that of the divergence-free one used for \eqref{eq:velocity_correction}.

\subsubsection{\textcolor{black}{RBF-HHD using least squares}}
To achieve a least squares solution for the RBF-HHD, reference points are introduced when constructing RBF kernels. A linear combination of kernels centered at the reference points, instead of the given data points, can approximate the given data with a continuous function.
The number of reference points should be fewer than that of the given data to ensure effective regression, i.e. avoid the danger of over-fitting.
The reference points do not have to be a subset of the centers, and they can be placed `arbitrarily' with specific preferences.
A quasi-uniform layout such as the Halton points for 2D is considered a desirable reference point placement strategy \citep{fasshauer2007meshfree, Larsson2023privaite}.
By introducing reference points, the matrices in \eqref{eq: div-free linear system} and \eqref{eq: curl-free linear system} are over-determined, and the coefficients ${\bf c}_j$ and $d_j$ are solved by least squares.

\subsection{PPE-HHD solver}
\label{sect: solvers ppe-hhd}
A PPE-HHD solver is adopted for the scaling law validations (see Sect.~\ref{sec: scaling law validation}).
This solver is used by \citet{mcclure2017instantaneous} to calculate the pressure gradient error field and here we adapt it to reconstruct pressure from pressure gradients.
The pressure gradients at boundaries are assumed to vanish.
Based on the HHD, the pressure gradients can be decomposed into a divergence-free and a curl-free part, as shown in \eqref{eq: poisson hhd}.
Incorporating the boundary conditions, i.e. the pressure gradients and their decomposed parts are zero at boundaries, the potentials ${\varphi}$ and ${\bf B}$ can be solved based on the Poisson equations:
\begin{equation}
\begin{split}
\left\{ \begin{matrix}
    \nabla^2 {\varphi}  &= - \nabla \cdot \nabla p \\ 
    \nabla^2 {\bf B}  &= - \textbf{curl} (\nabla p) \end{matrix}\right.
\end{split},
\label{eq: ppe-hhd}
\end{equation}
where the vector Laplacian is $\nabla^2 {\bf B} = 
\nabla (\nabla \cdot {\bf B}) - \textbf{curl} (\textbf{curl} (\bf B)) $; by setting $\bf B$ as divergence-free, we can obtain $- \textbf{curl} (\nabla p) = \nabla^2 {\bf B} - \nabla (\nabla \cdot {\bf B}) = \nabla^2 {\bf B}$.
$\varphi$ can be solved from \eqref{eq: ppe-hhd} using a second-order central difference Poisson solver.

\section*{Declarations}

\subsection*{Ethical Approval}
 
Not applicable. 

\subsection*{Competing interests}

The authors declare that they have no competing interests.
 
\subsection*{Authors' contributions}
Z.P. conceived the research.  J.W., Z.P., and L.L. provided the analysis for the scaling laws and error bounds. G.W., J.W., and J.M. provided the original RBF-HHD code, ODI code, and the PPE-HHD code, respectively. L.L. modified, integrated, and implemented the algorithms and performed the computations. 
All authors discussed the results and contributed to the manuscript.

\subsection*{Funding}
L.L. is partially supported by the Natural Sciences and Engineering Research Council of Canada (NSERC) Discovery Grant (RGPIN-2020-04486). G.B.W was partially supported by NSF grant DMS-1952674. J.P.W. was partially supported by NSF grant DMS-2206762.

\subsection*{Availability of data and materials}
Datasets and codes are available from the corresponding author, Z.P. on reasonable request.

\bibliographystyle{apalike}
\bibliography{PIV_Pressure_Lib}

\begin{thebibliography}{}

\bibitem[Alpaydin, 2020]{alpaydin2020introduction}
Alpaydin, E. (2020).
\newblock {\em Introduction to machine learning}.
\newblock MIT press.

\bibitem[Azijli and Dwight, 2015]{azijli2015solenoidal}
Azijli, I. and Dwight, R.~P. (2015).
\newblock {Solenoidal filtering of volumetric velocity measurements using
  Gaussian process regression}.
\newblock {\em Experiments in fluids}, 56(11):198.

\bibitem[Baur and K{\"o}ngeter, 1999]{baur1999piv}
Baur, T. and K{\"o}ngeter, J. (1999).
\newblock {PIV} with high temporal resolution for the determination of local
  pressure reductions from coherent turbulence phenomena.
\newblock In {\em {International Workshop on {PIV}'99- Santa Barbara, 3rd,
  Santa Barbara, CA}}, pages 101--106.

\bibitem[Bhatia et~al., 2012]{bhatia2012helmholtz}
Bhatia, H., Norgard, G., Pascucci, V., and Bremer, P.-T. (2012).
\newblock {The Helmholtz-Hodge decomposition—a survey}.
\newblock {\em IEEE Transactions on visualization and computer graphics},
  19(8):1386--1404.

\bibitem[Burkov, 2019]{burkov2019hundred}
Burkov, A. (2019).
\newblock {\em The hundred-page machine learning book}, volume~1.
\newblock Andriy Burkov Quebec City, QC, Canada.

\bibitem[Carr et~al., 2001]{carr2001reconstruction}
Carr, J.~C., Beatson, R.~K., Cherrie, J.~B., Mitchell, T.~J., Fright, W.~R.,
  McCallum, B.~C., and Evans, T.~R. (2001).
\newblock {Reconstruction and representation of 3D objects with radial basis
  functions}.
\newblock In {\em Proceedings of the 28th annual conference on Computer
  graphics and interactive techniques}, pages 67--76.

\bibitem[Charonko and Fratantonio, 2022]{Charonko2022Computational}
Charonko, J.~J. and Fratantonio, D. (2022).
\newblock {Computational optimization of omni-directional pressure integration
  schemes}.
\newblock 75th Annual Meeting of the Division of Fluid Dynamics.

\bibitem[Charonko et~al., 2010]{charonko2010assessment}
Charonko, J.~J., King, C.~V., Smith, B.~L., and Vlachos, P.~P. (2010).
\newblock {Assessment of pressure field calculations from particle image
  velocimetry measurements}.
\newblock {\em Measurement Science and Technology}, 21(10):105401.

\bibitem[Chattamvelli and Shanmugam, 2022]{chattamvelli2022continuous}
Chattamvelli, R. and Shanmugam, R. (2022).
\newblock {\em {Continuous Distributions in Engineering and the Applied
  Sciences--Part II}}.
\newblock Springer Nature.

\bibitem[Chorin et~al., 1990]{chorin1990mathematical}
Chorin, A.~J., Marsden, J.~E., and Marsden, J.~E. (1990).
\newblock {\em {A mathematical introduction to fluid mechanics}}, volume~3.
\newblock Springer.

\bibitem[Dabiri et~al., 2014]{dabiri2013algorithm}
Dabiri, J.~O., Bose, S., Gemmell, B.~J., Colin, S.~P., and Costello, J.~H.
  (2014).
\newblock {An algorithm to estimate unsteady and quasi-steady pressure fields
  from velocity field measurements}.
\newblock {\em Journal of Experimental Biology}, 217(3):331--336.

\bibitem[De~Kat and Van~Oudheusden, 2012]{de2012instantaneous}
De~Kat, R. and Van~Oudheusden, B. (2012).
\newblock {Instantaneous planar pressure determination from {PIV} in turbulent
  flow}.
\newblock {\em Experiments in fluids}, 52(5):1089--1106.

\bibitem[de~Silva et~al., 2013]{de2013minimization}
de~Silva, C.~M., Philip, J., and Marusic, I. (2013).
\newblock {Minimization of divergence error in volumetric velocity measurements
  and implications for turbulence statistics}.
\newblock {\em Experiments in fluids}, 54:1--17.

\bibitem[Drake et~al., 2022]{drake2022implicit}
Drake, K.~P., Fuselier, E.~J., and Wright, G.~B. (2022).
\newblock {Implicit surface reconstruction with a curl-free radial basis
  function partition of unity method}.
\newblock {\em SIAM J. Sci. Comput.}, 44(5):A3018--A3040.

\bibitem[Faiella et~al., 2021]{faiella2021error}
Faiella, M., Macmillan, C. G.~J., Whitehead, J.~P., and Pan, Z. (2021).
\newblock {Error propagation dynamics of velocimetry-based pressure field
  calculations (2): on the error profile}.
\newblock {\em Measurement Science and Technology}, 32(8):084005.

\bibitem[Fasshauer, 2007]{fasshauer2007meshfree}
Fasshauer, G.~E. (2007).
\newblock {\em {Meshfree approximation methods with MATLAB}}, volume~6.
\newblock World Scientific.

\bibitem[Fornberg and Flyer, 2015]{FFBook}
Fornberg, B. and Flyer, N. (2015).
\newblock {\em A {P}rimer on {R}adial {B}asis {F}unctions with {A}pplications
  to the {G}eosciences}.
\newblock SIAM, Philadelphia, USA.

\bibitem[Fornberg et~al., 2011]{fornberg2011stable}
Fornberg, B., Larsson, E., and Flyer, N. (2011).
\newblock {Stable computations with Gaussian radial basis functions}.
\newblock {\em SIAM Journal on Scientific Computing}, 33(2):869--892.

\bibitem[Fuselier et~al., 2016]{fuselier2016high}
Fuselier, E.~J., Shankar, V., and Wright, G.~B. (2016).
\newblock {A high-order radial basis function (RBF) Leray projection method for
  the solution of the incompressible unsteady Stokes equations}.
\newblock {\em Computers \& Fluids}, 128:41--52.

\bibitem[Fuselier and Wright, 2017]{fuselier2017radial}
Fuselier, E.~J. and Wright, G.~B. (2017).
\newblock {A radial basis function method for computing Helmholtz--Hodge
  decompositions}.
\newblock {\em IMA Journal of Numerical Analysis}, 37(2):774--797.

\bibitem[Gesemann et~al., 2016]{gesemann2016noisy}
Gesemann, S., Huhn, F., Schanz, D., and Schr{\"o}der, A. (2016).
\newblock {From noisy particle tracks to velocity, acceleration and pressure
  fields using B-splines and penalties}.
\newblock In {\em 18th international symposium on applications of laser and
  imaging techniques to fluid mechanics, Lisbon, Portugal}, volume~4.

\bibitem[Ghaemi et~al., 2012]{ghaemi2012piv}
Ghaemi, S., Ragni, D., and Scarano, F. (2012).
\newblock {PIV-based pressure fluctuations in the turbulent boundary layer}.
\newblock {\em Experiments in fluids}, 53(6):1823--1840.

\bibitem[Goushcha et~al., 2023]{goushcha2023modified}
Goushcha, O., Andreopoulos, Y., and Ganatos, P. (2023).
\newblock {A modified Green’s function approach to particle image velocimetry
  pressure estimates with an application to microenergy harvesters}.
\newblock {\em Archive of Applied Mechanics}, 93(3):1217--1239.

\bibitem[Gurka et~al., 1999]{gurka1999computation}
Gurka, R., Liberzon, A., Hefetz, D., Rubinstein, D., and Shavit, U. (1999).
\newblock {Computation of pressure distribution using PIV velocity data}.
\newblock In {\em Workshop on particle image velocimetry}, volume~2, pages
  1--6.

\bibitem[Harlander et~al., 2014]{harlander2014orthogonal}
Harlander, U., von Larcher, T., Wright, G.~B., Hoff, M., Alexandrov, K., and
  Egbers, C. (2014).
\newblock Orthogonal decomposition methods to analyze {PIV}, {LDA} and
  thermography data of a thermally driven rotating annulus laboratory
  experiment.
\newblock In von Larcher, T. and Williams, P.~D., editors, {\em {Modelling
  Atmospheric and Oceanic flows: insights from laboratory experiments and
  numerical simulations}}, pages 315--336. American Geophysical Union,
  Washington D.C.

\bibitem[Huhn et~al., 2016]{huhn2016fft}
Huhn, F., Schanz, D., Gesemann, S., and Schr{\"o}der, A. (2016).
\newblock {FFT integration of instantaneous 3D pressure gradient fields
  measured by Lagrangian particle tracking in turbulent flows}.
\newblock {\em Experiments in Fluids}, 57(9):151.

\bibitem[Iserles, 2009]{iserles2009first}
Iserles, A. (2009).
\newblock {\em {A first course in the numerical analysis of differential
  equations}}.
\newblock Cambridge university press.

\bibitem[Larsson, 2023]{Larsson2023privaite}
Larsson, E. (2023).
\newblock Private Communication.

\bibitem[Larsson et~al., 2013]{larsson2013stable}
Larsson, E., Lehto, E., Heryudono, A., and Fornberg, B. (2013).
\newblock {Stable computation of differentiation matrices and scattered node
  stencils based on Gaussian radial basis functions}.
\newblock {\em SIAM Journal on Scientific Computing}, 35(4):A2096--A2119.

\bibitem[Lee et~al., 2022]{lee2022effect}
Lee, H., Park, H.~J., Kim, M., Han, J., and Hwang, W. (2022).
\newblock {The effect of perspective error on 2D PIV Measurements of
  homogeneous isotropic turbulence}.
\newblock {\em Experiments in Fluids}, 63(8):122.

\bibitem[Li and Pan, 2024]{li2024three}
Li, L. and Pan, Z. (2024).
\newblock {Three-dimensional time-resolved Lagrangian flow field reconstruction
  based on constrained least squares and stable radial basis function}.
\newblock {\em Experiments in Fluids}, 65(4):57.

\bibitem[Lin and Xu, 2023]{lin2023Divergence}
Lin, Y. and Xu, H. (2023).
\newblock {Divergence–curl correction for pressure field reconstruction from
  acceleration in turbulent flows}.
\newblock {\em Experiments in Fluids}, 64(137).

\bibitem[Liu and Katz, 2006]{liu2006instantaneous}
Liu, X. and Katz, J. (2006).
\newblock {Instantaneous pressure and material acceleration measurements using
  a four-exposure {PIV} system}.
\newblock {\em Experiments in Fluids}, 41(2):227--240.

\bibitem[Liu and Moreto, 2020]{liu2020error}
Liu, X. and Moreto, J.~R. (2020).
\newblock {Error propagation from the PIV-based pressure gradient to the
  integrated pressure by the omnidirectional integration method}.
\newblock {\em Measurement Science and Technology}, 31(5):055301.

\bibitem[Liu et~al., 2016]{liu2016instantaneous}
Liu, X., Moreto, J.~R., and Siddle-Mitchell, S. (2016).
\newblock {Instantaneous pressure reconstruction from measured pressure
  gradient using rotating parallel ray method}.
\newblock In {\em 54th AIAA Aerospace Sciences Meeting}, page 1049.

\bibitem[Mac{\^e}do et~al., 2011]{macedo2011hermite}
Mac{\^e}do, I., Gois, J.~P., and Velho, L. (2011).
\newblock {Hermite radial basis functions implicits}.
\newblock {\em Computer Graphics Forum}, 30(1):27--42.

\bibitem[McClure and Yarusevych, 2017]{mcclure2017instantaneous}
McClure, J. and Yarusevych, S. (2017).
\newblock Instantaneous {PIV/PTV-based} pressure gradient estimation: a
  framework for error analysis and correction.
\newblock {\em Experiments in Fluids}, 58:1--18.

\bibitem[McClure and Yarusevych, 2019]{mcclure2019generalized}
McClure, J. and Yarusevych, S. (2019).
\newblock Generalized framework for {PIV-based} pressure gradient error field
  determination and correction.
\newblock {\em Measurement Science and Technology}, 30(8):084005.

\bibitem[Nie et~al., 2022]{Nie2022Error}
Nie, M., Whitehead, J.~P., Richards, G., Smith, B.~L., and Pan, Z. (2022).
\newblock {Error propagation dynamics of PIV-based pressure field calculation
  (3): what is the minimum resolvable pressure in a reconstructed field?}
\newblock {\em Experiments in Fluids}, 63(11):168.

\bibitem[Pan et~al., 2016]{Pan2016Error1}
Pan, Z., Whitehead, J., Thomson, S., and Truscott, T. (2016).
\newblock {Error propagation dynamics of PIV-based pressure field calculations:
  How well does the pressure Poisson solver perform inherently?}
\newblock {\em Measurement Science and Technology}, 27(8):084012.

\bibitem[Panton, 2006]{panton2006incompressible}
Panton, R.~L. (2006).
\newblock {\em {Incompressible flow}}.
\newblock John Wiley \& Sons.

\bibitem[Pirnia et~al., 2020]{pirnia2020estimating}
Pirnia, A., McClure, J., Peterson, S., Helenbrook, B., and Erath, B. (2020).
\newblock {Estimating pressure fields from planar velocity data around immersed
  bodies: a finite element approach}.
\newblock {\em Experiments in Fluids}, 61(2):1--16.

\bibitem[Pryce et~al., 2024a]{prycerevisit}
Pryce, C., Li, L., and Pan, Z. (2024a).
\newblock {Revisit Liu \& Katz (2006) and Zigunov \& Charonko (2024): On the
  Equivalency of Omni-directional Integration and Pressure Poisson Equation}.
\newblock In {\em 21th international symposium on applications of laser and
  imaging techniques to fluid mechanics, Lisbon, Portugal}.

\bibitem[Pryce et~al., 2024b]{Pryce2024revisit}
Pryce, C., Li, L., and Pan, Z. (2024b).
\newblock {Revisit Liu and Katz (2006) and Zigunov and Charonko (2024b), Part
  (I): on the Equivalence of the Omnidirectional Integration and the Pressure
  Poisson Equation}.
\newblock {\em arXiv preprint arXiv:2411.02583}.

\bibitem[Pryce et~al., 2024c]{pryce2024simple}
Pryce, C., Li, L., Whitehead, J.~P., and Pan, Z. (2024c).
\newblock {A simple boundary condition regularization strategy for image
  velocimetry-based pressure field reconstruction}.
\newblock {\em Experiments in Fluids}, 65(6):1--6.

\bibitem[Raffel et~al., 2018]{Raffel2018}
Raffel, M., Willert, C.~E., Scarano, F., K{\"a}hler, C.~J., Wereley, S.~T., and
  Kompenhans, J. (2018).
\newblock {\em Techniques for 3D-PIV}, pages 309--365.
\newblock Springer International Publishing, Cham.

\bibitem[Ratz and Mendez, 2024]{ratz2024meshless}
Ratz, M. and Mendez, M.~A. (2024).
\newblock {A meshless and binless approach to compute statistics in 3D ensemble
  PTV}.
\newblock {\em Experiments in Fluids}, 65(9):142.

\bibitem[Rice, 1944]{rice1944mathematical}
Rice, S.~O. (1944).
\newblock {Mathematical analysis of random noise}.
\newblock {\em The Bell System Technical Journal}, 23(3):282--332.

\bibitem[Schiavazzi et~al., 2017]{schiavazzi2017effect}
Schiavazzi, D.~E., Nemes, A., Schmitter, S., and Coletti, F. (2017).
\newblock {The effect of velocity filtering in pressure estimation}.
\newblock {\em Exp Fluids}, 58(5):50.

\bibitem[Schr{\"a}der and Wendland, 2011]{schrader2011high}
Schr{\"a}der, D. and Wendland, H. (2011).
\newblock {A high-order, analytically divergence-free discretization method for
  Darcy’s problem}.
\newblock {\em Mathematics of computation}, 80(273):263--277.

\bibitem[Schwabe, 1935]{schwabe1935druckermittlung}
Schwabe, M. (1935).
\newblock {\"U}ber druckermittlung in der nichtstation{\"a}ren ebenen
  str{\"o}mung.
\newblock {\em Ingenieur-Archiv}, 6(1):34--50.

\bibitem[Sperotto et~al., 2022]{sperotto2022meshless}
Sperotto, P., Pieraccini, S., and Mendez, M.~A. (2022).
\newblock {A meshless method to compute pressure fields from image
  velocimetry}.
\newblock {\em Measurement Science and Technology}, 33(9):094005.

\bibitem[Van~der Kindere et~al., 2019]{van2019pressure}
Van~der Kindere, J., Laskari, A., Ganapathisubramani, B., and De~Kat, R.
  (2019).
\newblock {Pressure from 2D snapshot PIV}.
\newblock {\em Experiments in fluids}, 60:1--18.

\bibitem[Van~Gent et~al., 2017]{van2017comparative}
Van~Gent, P., Michaelis, D., Van~Oudheusden, B., Weiss, P.-{\'E}., de~Kat, R.,
  Laskari, A., Jeon, Y.~J., David, L., Schanz, D., Huhn, F., et~al. (2017).
\newblock {Comparative assessment of pressure field reconstructions from
  particle image velocimetry measurements and Lagrangian particle tracking}.
\newblock {\em Experiments in Fluids}, 58(4):1--23.

\bibitem[Van~Oudheusden, 2013]{van2013piv}
Van~Oudheusden, B. (2013).
\newblock {PIV}-based pressure measurement.
\newblock {\em Measurement Science and Technology}, 24(3):032001.

\bibitem[Wang et~al., 2017]{wang2017weighted}
Wang, C., Gao, Q., Wei, R., Li, T., and Wang, J. (2017).
\newblock {Weighted divergence correction scheme and its fast implementation}.
\newblock {\em Experiments in Fluids}, 58:1--14.

\bibitem[Wang, 2019]{wang2019impact}
Wang, J. (2019).
\newblock {\em {Impact of Pressure on Deformation of a Compliant Wall in a
  Turbulent Boundary Layer}}.
\newblock PhD thesis, The Johns Hopkins University.

\bibitem[Wang, 2023]{Wang2023private}
Wang, J. (2023).
\newblock Private Communication.

\bibitem[Wang et~al., 2019]{wang2019gpu}
Wang, J., Zhang, C., and Katz, J. (2019).
\newblock {GPU-based, parallel-line, omni-directional integration of measured
  pressure gradient field to obtain the 3D pressure distribution}.
\newblock {\em Experiments in Fluids}, 60:1--24.

\bibitem[Wang and Liu, 2023]{wang2023green}
Wang, Q. and Liu, X. (2023).
\newblock {Green's function integral method for pressure reconstruction from
  measured pressure gradient and the interpretation of omnidirectional
  integration}.
\newblock {\em Physics of Fluids}, 35(7).

\bibitem[Wang et~al., 2016]{wang2016irrotation}
Wang, Z., Gao, Q., Wang, C., Wei, R., and Wang, J. (2016).
\newblock {An irrotation correction on pressure gradient and orthogonal-path
  integration for PIV-based pressure reconstruction}.
\newblock {\em Experiments in Fluids}, 57:1--16.

\bibitem[Weller et~al., 1998]{weller1998tensorial}
Weller, H.~G., Tabor, G., Jasak, H., and Fureby, C. (1998).
\newblock {A tensorial approach to computational continuum mechanics using
  object-oriented techniques}.
\newblock {\em Computers in physics}, 12(6):620--631.

\bibitem[Wendland, 2009]{wendland2009divergence}
Wendland, H. (2009).
\newblock {Divergence-free kernel methods for approximating the Stokes
  problem}.
\newblock {\em SIAM Journal on Numerical Analysis}, 47(4):3158--3179.

\bibitem[Zhou et~al., 2023]{zhou2023stochastic}
Zhou, K., Li, J., Hong, J., and Grauer, S.~J. (2023).
\newblock {Stochastic particle advection velocimetry (SPAV): theory,
  simulations, and proof-of-concept experiments}.
\newblock {\em Measurement Science and Technology}, 34(6):065302.

\bibitem[Zigunov and Charonko, 2024a]{zigunov2024fast}
Zigunov, F. and Charonko, J.~J. (2024a).
\newblock {A fast, matrix-based method to perform omnidirectional pressure
  integration}.
\newblock {\em Measurement Science and Technology}, 35(6):065302.

\bibitem[Zigunov and Charonko, 2024b]{zigunov2024one}
Zigunov, F. and Charonko, J.~J. (2024b).
\newblock {One-shot omnidirectional pressure integration through matrix
  inversion}.
\newblock {\em Measurement Science and Technology}, 35(12):125301.

\end{thebibliography}

\end{document}